\documentclass[preprint,amssymb,aps,prb,superscriptaddress]{revtex4-1}  %uncomment this line to have single column
\usepackage{amsmath}
\usepackage{amssymb}
\usepackage{bm}
\usepackage{epsfig}
\usepackage{graphicx}
\usepackage{color}
\usepackage{hhline}

\usepackage{color}

\newcommand{\inlinecite}[1]{[\onlinecite{#1}]}

\def\be{\begin{equation}}
\def\ee{\end{equation}}
\def\bea{\begin{eqnarray}}
\def\eea{\end{eqnarray}}

\begin{document}
\newcount\timehh  \newcount\timemm
\timehh=\time \divide\timehh by 60
\timemm=\time
\count255=\timehh\multiply\count255 by -60 \advance\timemm by \count255

\title{Electronic states and optical properties of PbSe nanorods and nanowires}

\author{A. C. Bartnik}
\affiliation{Applied Physics Department, Cornell University, Ithaca, NY 14853}
\author{Al. L. Efros}
\affiliation{Naval Research Laboratory, Washington, DC 20375}
\author{W.-K. Koh}
\affiliation{Department of Chemistry, University of Pennsylvania, Philadelphia, Pennsylvania 19104}
\author{C. B. Murray}
\affiliation{Department of Chemistry, University of Pennsylvania, Philadelphia, Pennsylvania 19104}
\affiliation{Department of Materials Science and Engineering, University of Pennsylvania, Philadelphia, Pennsylvania 19104}
\author{F. W. Wise}
\affiliation{Applied Physics Department, Cornell University, Ithaca, NY 14853}

\begin{abstract}
A theory of the electronic structure and excitonic absorption spectra of PbS and PbSe nanowires and nanorods in the framework of a four–-band effective mass model is presented. Calculations conducted for PbSe show that dielectric contrast dramatically strengthens the exciton binding in narrow nanowires and nanorods.  However, the self-interaction energies of the electron and hole nearly cancel the Coulomb binding, and as a result the optical absorption spectra are practically unaffected by the strong dielectric contrast between PbSe and the surrounding medium.  Measurements of the size-dependent absorption spectra of colloidal PbSe nanorods are also presented. Using room-temperature energy-band parameters extracted from the optical spectra of spherical PbSe nanocrystals, the theory provides good quantitative agreement with the measured spectra.
\end{abstract}

\maketitle

\section{Introduction}

Solution-based chemical synthesis of semiconductor nanostructures has allowed tremendous flexibility in crystal morphology.  After much work on zero-dimensional (0D) nanocrystals (NCs), attention is shifting to one-dimensional (1D) nanorods (NRs) and nanowires (NWs) \cite{kuno, jun_angewandte, burda, scher_alivisatos} and the variation of material properties in the transition from 0D to 1D. The electronic structure of these crystals is the foundation for understanding their properties. Previously, the electronic structure of 1D nanocrystals has been modeled using a variety of methods, including effective-mass theories based on ${\bm k} \cdot {\bm p}$ Hamiltonians \cite{muljarov_masumoto,muljarov_masumoto_2,shabaev_efros, bockelmann_kp, sun_dft_kp}, pseudopotential techniques \cite{zunger_pseudo_1, zunger_pseudo_2, franc_zunger_pseudo}, tight binding models \cite{persson_tb_1, persson_tb_2, persson_tb_3, delerue_tightbind, charlier_tb_dft}, and density functional theory \cite{serra_dft, landman_dft, prezhdo_dft, sun_dft_kp, charlier_tb_dft, bruno_dft}. The relaxation of confinement in going from 0D to 1D goes hand--in--hand with an increase in the importance of Coulomb effects mediated through the nanocrystal's dielectric environment \cite{bruscomment}.

Lead--salt (PbS, PbSe, PbTe) nanocrystals offer unique advantages to study the interplay of these two effects. Their large exciton Bohr radii places them at the limit of strong confinement, while their large dielectric constants coupled with their mirror--like electron and hole spectra substantially reduce the Coulomb interaction in spherical quantum dots \cite{wise_acr, kang_wise}. However, in a 1D structure the Coulomb interaction can act primarily through the host medium, so it will not be screened as effectively as in 0D \cite{shabaev_efros}. Thus, the lead salts provide a unique system to study the transition from strong confinement to strong Coulomb binding as the length of the nanocrystal changes.

Within ${\bm k} \cdot {\bm p}$ theory, the general treatment of the optical properties of NWs and NRs surrounded by media with small dielectric constant was developed in Refs. \inlinecite{muljarov_masumoto,muljarov_masumoto_2,shabaev_efros}. A type of adiabatic approximation naturally separates the calculation into pieces.  In recognition of strong confinement perpendicular to the NR or NW axis, one first calculates the 1D subband energies and wavefunctions, while neglecting the Coulomb interaction.  Next, using these wave functions of transverse electron and hole motion, one can calculate the longitudinal motion of the exciton, including corrections from image forces in the surrounding medium.  To do that, the three-dimensional Coulomb potential is averaged to a one-dimensional Coulomb interaction between the electron and hole along the NW or NR axis. Using this potential, the spectra of 1D excitons and their transition oscillator strengths are found. Finally, in NRs one should find the spectrum of the exciton center of mass motion, in order to include this additional effect of confinement. The main aspects of this framework were performed for lead--salt nanowires recently by Rupasov \cite{rupasov_1}, although approximations to the simplified band structure used in that paper preclude the description of real experimental results.

In this paper we present calculations of the 1D subband energy spectra of lead--salt nanowires with arbitrary axis orientation, taking into account the multi-valley structure and accurate electron and hole energy-level dispersions in these semiconductors. For PbSe NWs with axis along the $\langle100\rangle$ direction, we calculate the spectra of 1D excitons including self-interaction corrections. Surprisingly, the calculations show that although the binding energy of excitons in the smallest NWs reaches 350 meV, the optical transition energies are not affected by the small dielectric constant of the surrounding medium and are almost identical to the transitions between non-interacting electron and hole subbands. The cancelation of the exciton binding energy and the self-interaction corrections to the electron and hole levels is a consequence of the almost mirror symmetry of the conduction and valence bands of PbSe. The theoretical results agree well with the measured absorption spectra of $\langle100\rangle$ PbSe NRs.

The paper is organized as follows. In Section \ref{fourband} we will describe the Hamiltonian governing the 1D nanowire system, with solutions in Section \ref{nanowire}. In Section \ref{coulomb} we present the effects of dielectric confinement and Coulomb forces on the 1D exciton, with 1D wavefunction solutions in Section \ref{1Dexc}. Experimental data and comparison with theory are presented in Section \ref{experiment}, followed by discussion and conclusion.

\section{\label{fourband}Four band effective mass model}

PbS, PbSe, and PbTe are direct-gap semiconductors, with
extrema of the conduction and valence bands at the $L$ points in the Brillouin zone. The energy bands near the $L$ point can be well-described within the four--band $\mathbf{k \cdot p}$  model \cite{dimmock_wright, mitchell_wallis}. This model takes into account the direct interaction between the nearest conduction and valence bands, as well as the contributions of the remote bands to the electron and hole effective masses. Following Ref. \inlinecite{kang_wise}, we use the multiband effective mass approximation and expand the full wave functions inside the nanorod as
\begin{eqnarray}
{\Phi (\mbox{\boldmath $ r$})} &=& \sum_{\mu = \pm 1/2}
\Psi^{c}_\mu(\mbox{\boldmath $ r$}) |L_{6,\mu}^{-}\rangle
 \, + \sum_{\mu = \pm 1/2}\Psi^{v}_\mu(\mbox{\boldmath $ r$})
|L_{6,\mu}^{+}\rangle
 \,
,\label{psi}
\end{eqnarray}
where $|L_{6,\mu}^{-}\rangle $ and  $|L_{6,\mu}^{+}\rangle $ are the Bloch functions of the conduction band and valence band edge, respectively, at the L--point. The upper sign ``$\pm$'' in the notation reflects the invariance of these functions with respect to the operation of spatial inversion. The smooth functions $\Psi^{c}_{\pm 1/2}(\mbox{\boldmath $r$})$ and $\Psi^{v}_{\pm 1/2}(\mbox{\boldmath $r$})$ are the components of the conduction band and valence band spinor envelopes, respectively:
\begin{eqnarray}
\Psi^c=\left( \begin{array}{c} \Psi^c_{1/2}
\\ \Psi^c_{-1/2} \end{array} \right) \, , \qquad \Psi^v=\left(
\begin{array}{c} \Psi^v_{1/2}
\\ \Psi^v_{-1/2} \end{array} \right).
\end{eqnarray}
 The bi--spinor
envelope function $\Psi=\left( \begin{array}{c} \Psi^c
\\ \Psi^v \end{array} \right)$ is the solution of the Schr\"{o}dinger
equation
$\hat H(\hat {\bm p}) \Psi = E \Psi$, where $ \mbox{\boldmath $\hat  p$}=
\hbar \hat {\bm k}= - i \hbar \mbox{\boldmath $\nabla$}$ is the momentum operator,  and the Hamiltonian $\hat H( \hat {\bm p})$ of Ref. \inlinecite{kang_wise}
can be written in compact form as
\begin{eqnarray}
   \hat{H}( \hat{\mbox{\boldmath $  p$}})=
   \left(  \begin{array}{cc} \left(\dfrac{E_g}{2}+
\dfrac{\hat{p}^{2}_z}{2m_l^{-}}  +\dfrac{\hat{p}^{2}_\perp}{2m_t^{-}}  \right)  \hat U_2 &
    \dfrac{P_l}{m_0} \hat{p}_z \hat{\sigma}_z +  \dfrac{P_t}{m_0} \left( \hat{p}_\perp
\hat{\sigma}_\perp\right)   \\
 \dfrac{P_l}{m_0} \hat{p}_z \hat{\sigma}_z +  \dfrac{P_t}{m_0} \left( \hat{p}_\perp
\hat{\sigma}_\perp\right)  &
- \left(\dfrac{E_g}{2}+ \dfrac{\hat{p}^{2}_z}{2m_l^{+}}
+\dfrac{\hat{p}^{2}_\perp}{2m_t^{+}}  \right)   \hat U_2
\end{array} \right)
 \, \,  .  \label{hfull}
\end{eqnarray}
 In Eq. \eqref{hfull} $\hat U_2$ is the $2\times 2$ unit matrix,  $ \hat {\bm \sigma}= \{ \hat \sigma_x , \hat \sigma_y ,  \hat \sigma_z \}$  are the Pauli matrices that act on the spinor components of the wave functions ($\mu= \pm 1/2$), $E_g$ is the bulk energy gap, $E$ is the electron or hole energy measured from the middle of the gap, $m_0$ is the free electron mass, $\hat{p}_\perp^2=\hat{p}_x^2+\hat{p}_y^2$, $( \hat{p}_\perp \hat{\sigma}_\perp)=\hat{p}_x \hat{\sigma}_x + \hat{p}_y \hat{\sigma}_y$,  $P_t$ and $P_l$ are the transverse and longitudinal momentum matrix elements taken between the conduction and valence band edge Bloch functions \cite{kang_wise}, and $m_t^\pm$ and $m_l^\pm$ are the remote-band contribution to the transverse and longitudinal band edge effective masses, respectively.
 For electrons and holes, these band edge effective masses can be expressed as $m^e_{l,t}=[1/m^-_{l,t}+2 P^2_{l,t}/ m_0^2 E_g]^{-1}$ and $m^h_{l,t}=[1/m^+_{l,t}+2 P^2_{l,t}/m_0^2 E_g]^{-1}$, respectively. In each valley, the $z$ axis in Eq. \eqref{hfull} is directed toward the L--point of the Brillouin zone, e.g. along the $\langle 111\rangle$ direction of the cubic lattice. As a result, for each of the four valleys, the $z$ axis will point in different directions.

Although the Hamiltonian of Eq. \eqref{hfull} has cylindrical symmetry with respect to, e.g., the $\langle 111\rangle$ crystallographic direction, this direction may not coincide with the NR growth direction. For a description of NR electronic and optical properties it is convenient to use coordinates connected with the latter direction instead, even though the cylindrical symmetry of the Hamiltonian is generally broken.  In PbS and PbSe, the small anisotropy of conduction and valence bands allows us to treat deviations from cylindrical symmetry perturbatively.
 The Hamiltonian \eqref{hfull} can be written  $\hat H
= \hat H_0 + \hat H_{an}$, where the cylindrically-symmetric part $\hat H_0$ is
\begin{eqnarray}
   \hat{H_0}( \hat{\mbox{\boldmath $  p$}}) & = &
   \left(  \begin{array}{cc} \left(\dfrac{E_g}{2}+
\dfrac{\hat{p}^{2}_z}{2m_z^{-}}  +\dfrac{\hat{p}^{2}_\perp}{2m_\perp^{-}}  \right)  \hat U &
    \dfrac{P_z}{m_0} \hat{p}_z \hat{\sigma}_z +  \dfrac{P_\perp}{m_0} \left( \hat{p}_\perp
\hat{\sigma}_\perp\right)   \\
 \dfrac{P_z}{m_0} \hat{p}_z \hat{\sigma}_z +  \dfrac{P_\perp}{m_0} \left( \hat{p}_\perp
\hat{\sigma}_\perp\right)  &
- \left(\dfrac{E_g}{2}+ \dfrac{\hat{p}^{2}_z}{2m_z^{+}}
+\dfrac{\hat{p}^{2}_\perp}{2m_\perp^{+}}  \right)   \hat U
\end{array} \right)
 \, \,  .  \label{hcyl}
\end{eqnarray}
The modified band parameters are
\begin{eqnarray}
 P_\perp & = & \frac{P_t}{2}  (1 + \cos^2 \theta) + \frac{ P_l}{2} \sin^2\theta \qquad \qquad P_z  =  P_t \sin^2 \theta + P_l \cos^2 \theta \label{modifiedparam}\\
 \frac{1}{m_\perp^\pm} & = & \frac{1}{2m_t^\pm} (1 + \cos^2 \theta) + \frac{1}{2m_l^\pm} \sin^2\theta \qquad \quad \frac{1}{m_z^\pm} = \frac{1}{m_t^\pm} \sin^2\theta + \frac{1}{m_l^\pm} \cos^2 \theta
\end{eqnarray}
where $\theta$ is the angle between the growth axis and the $\langle 111\rangle$ direction. The anisotropic part of the Hamiltonian is given in Appendix \ref{pertappendix}. Note that Eq. \eqref{hcyl} has a form identical to Eq. \eqref{hfull}, but the $z$ axis is now directed along the growth axis. For arbitrary orientation of the growth direction, there will be four different angles $\theta$ for each of the four valleys, and therefore four different sets of modified band parameters defined in Eq. \eqref{modifiedparam}. As a result, each valley will have unique electronic structure.

The energy spectra associated with the different valleys become degenerate when the growth direction leads to identical values of $\theta$ for them.  The highest degree of degeneracy is reached when the growth direction is along the $\langle 100 \rangle$ crystal axis. In this case all four valleys have the same $\theta$; $\cos^2\theta = 1/3$, which results in $P_\perp = P_z$ and $m_\perp = m_z$ in Eq. \eqref{hcyl}.  All of the spectra are degenerate.

The anisotropic part $\hat{H}_{an}$  of the full Hamiltonian can be considered as a perturbation  if  $|P_l-P_t|\ll P_l+P_t$ and $|1/m_l^{\pm}-1/m_t^{\pm}|\ll 1/m_l^{\pm}+1/m_t^{\pm}$.   The first-order corrections to the solutions of $\hat{H}_{0}$ caused by $\hat{H}_{an}$ vanish in the 2-fold Kramers-degenerate subspace at each energy level. As a result, only second-order perturbation theory gives corrections to the energy levels. We will neglect these corrections from this point on, although an example higher-order calculation appears in Appendix \ref{pertappendix}.

\section{\label{nanowire}Energy spectra of electrons and holes in PbSe Nanowires}
The first step in our modeling process is to find the energy spectra of 1D subbands of infinitely-long cylindrical nanowires, temporarily ignoring the  Coulomb interaction. The cylindrical symmetry of the Hamiltonian of Eq. \eqref{hcyl} allows the solutions to take the form
\begin{equation}
\Psi^n(k_z) = \begin{pmatrix}
\mathcal{R}_1^n(\rho) \exp(i (n-1/2)\phi) \\
i \mathcal{R}_2^n(\rho) \exp(i (n+1/2)\phi) \\
\mathcal{R}_3^n(\rho) \exp(i (n-1/2)\phi) \\
i \mathcal{R}_4^n(\rho) \exp(i (n+1/2)\phi)
\end{pmatrix} \exp(i k_z z) ~ ,
\label{angular}
\end{equation}
where $\phi$ is the azimuthal angle,  $n=\pm1/2,\pm3/2,\pm5/2, ...$ is the total angular momentum projection on the nanowire axes defined by the operator $\hat{J}_z = -i\partial/\partial \phi + \hat{S}_z$, $\hbar k_z$ is the momentum along the nanowire $z$ axis, and $\rho=\sqrt{x^2+y^2}$ is the radial coordinate in the plane perpendicular to the NW axis. The chosen phase of each component of the function $\Psi^n(k_z)$ allows the radial functions $\mathcal{R}_i^n(\rho)$ to be pure real. Substitution of Eq. \eqref{angular} into Eq. \eqref{hcyl} yields the system of differential equations that defines these functions:
\bea
 \left(\alpha_-  + \frac{\hbar^2}{2 m_\perp^-}\Delta_{n-1/2} \right)\mathcal{R}_1^n(\rho) + \frac{\hbar k_z P_z}{m_0} \mathcal{R}_3^n(\rho) + \frac{\hbar P_\perp}{m_0} \hat{D}^-_{n+1/2} \mathcal{R}_4^n(\rho)& = &0~,\nonumber \\
\left(\alpha_-  + \frac{\hbar^2}{2 m_\perp^-} \Delta_{n+1/2}\right)\mathcal{R}_2^n(\rho) + \frac{\hbar P_\perp}{m_0} \hat{D}^+_{n-1/2} \mathcal{R}_3^n(\rho) - \frac{\hbar k_z P_z}{m_0} \mathcal{R}_4^n(\rho) & = & 0~,\nonumber\\
- \frac{\hbar k_z P_z}{m_0} \mathcal{R}_1^n(\rho) - \frac{\hbar P_\perp}{m_0} \hat{D}^-_{n+1/2}\mathcal{R}_2^n(\rho)+\left(\alpha_+  + \frac{\hbar^2}{2 m_\perp^-} \Delta_{n-1/2}\right)\mathcal{R}_3^n(\rho) & = &0~,\nonumber \\
 - \frac{\hbar P_\perp}{m_0} \hat{D}^+_{n-1/2} \mathcal{R}_1^n(\rho)+ \frac{\hbar k_z P_z}{m_0} \mathcal{R}_2^n(\rho)+\left(\alpha_+ + \frac{\hbar^2}{2 m_\perp^-} \Delta_{n+1/2}\right)\mathcal{R}_4^n(\rho) & = & 0~,
\label{diffeqs}
\eea
where $\alpha_\pm=E_g/2 \pm E + \hbar^2 k_z^2/(2 m_z^\pm)$. The differential operators
\be
\hat{D}^\pm_m =  \mp \frac{\partial}{\partial \rho} + \frac{m}{\rho}
\ee
are the raising and lowering operators $\hat{D}^\pm_m J_m(k \rho) = k J_{m \pm 1} (k \rho)$ for the Bessel functions $J_m(k \rho)$ with integer index, and the operator $\Delta_m=\hat{D}^-_{m+1}\hat{D}^+_m=-(1/\rho)(\partial/\partial\rho) \rho(\partial/\partial\rho)+m^2/\rho^2$.

It is easy to check using the raising and lowering properties of the $\hat{D}^\pm_m$ operators that the radial eigenfunctions of Eqs. \eqref{diffeqs} should take the form
\be
\begin{pmatrix}
\mathcal{R}_1^n(\rho)  \\
\mathcal{R}_2^n(\rho)  \\
\mathcal{R}_3^n(\rho)  \\
\mathcal{R}_4^n(\rho)
\end{pmatrix} =
\begin{pmatrix}
C_1J_{n-1/2}(k_\rho \rho) \\
C_2J_{n+1/2}(k_\rho \rho) \\
C_3J_{n-1/2}(k_\rho \rho) \\
C_4J_{n+1/2}(k_\rho \rho)
\end{pmatrix}.
\label{radsol}
\ee
Substitution of this into Eqs. \eqref{diffeqs} yields a 4x4 system of linear equations for the coefficients $C_{1,2,3,4}$. Setting the determinant of this system to zero produces the relation between the quasi-momentum $k_\rho$  and the energy of electrons or holes $E$:
\be
\hbar^2 k_\rho^2 = - \alpha(E) \pm \sqrt{\alpha(E)^2 + \beta(E)}~,
\label{k2}
\ee
where
\bea
\alpha(E) & = & m_\perp^+ \left(E + \frac{\hbar^2 k_z^2}{2 m_z^+} + \frac{E_g}{2} \right) - m_\perp^- \left(E - \frac{\hbar^2 k_z^2}{2 m_z^-} - \frac{E_g}{2} \right) + m_\perp^- m_\perp^+ \frac{2 P_\perp^2}{m^2} \nonumber\\
\beta(E) & = & 4 m_\perp^+ m_\perp^- \left(E + \frac{\hbar^2 k_z^2}{2 m_z^+} + \frac{E_g}{2}  \right) \left(E  - \frac{\hbar^2 k_z^2}{2 m_z^-} - \frac{E_g}{2}\right) - 4  \frac{ m_\perp^- m_\perp^+ }{m^2} P_z^2 \hbar^2 k_z^2~.
\eea
From Eq. \eqref{k2} it is clear that $k^2_\rho$ can be positive or negative.  The negative value of $k^2_\rho$ results in an imaginary $k_\rho=i\lambda_\rho$, with $\lambda_\rho$  defined by Eq. \eqref{k2} as $\hbar^2 \lambda_\rho^2 = \alpha(E) + \sqrt{\alpha(E)^2 + \beta(E)}$. The complex arguments in Eq. \eqref{radsol} are then simplified by replacing the Bessel functions $J_m(i\lambda_\rho\rho)$ with the modified Bessel functions $I_m(\lambda_\rho\rho)$ using the relationship  $J_m(i\lambda_\rho\rho)=i^mI_m(\lambda_\rho\rho)$.  For each value of $k^2_\rho$, there are two independent solutions of the 4x4 linear system for the coefficients $C_{1,2,3,4}$.  These two solutions can be chosen such that either $C_3=0$ or $C_4=0$, which allows the remaining coefficients $C_i$ to be found. Taking into account the positive and negative value of $k^2_\rho$, there are four independent solutions for each energy $E$.

The energy spectrum is determined by the boundary conditions at the NW surface. The boundary conditions are defined on all four components of the wave function, which inside of the NW  can be always written as a linear combination of the four degenerate solutions discussed above
\begin{eqnarray}
\begin{pmatrix}
\mathcal{R}_1^n(\rho,k_z)  \\
\mathcal{R}_2^n(\rho,k_z)  \\
\mathcal{R}_3^n(\rho,k_z)  \\
\mathcal{R}_4^n(\rho,k_z)
\end{pmatrix} & = &
a \begin{pmatrix}
k_\rho P_\perp J_{n-1/2}(k_\rho \rho) \\
-k_z P_z J_{n+1/2}(k_\rho \rho) \\
0  \\
\Gamma_k J_{n+1/2}(k_\rho \rho)
\end{pmatrix} +
b \begin{pmatrix}
k_z P_z J_{n-1/2}(k_\rho \rho) \\
k_\rho P_\perp J_{n+1/2}(k_\rho \rho) \\
\Gamma_k  J_{n-1/2}(k_\rho \rho) \\
0
\end{pmatrix} + \nonumber\\
& & \qquad +
c \begin{pmatrix}
\lambda_\rho P_\perp I_{n-1/2}(\lambda_\rho \rho) \\
-k_z P_z I_{n+1/2}(\lambda_\rho \rho) \\
0  \\
\Gamma_\lambda I_{n+1/2}(\lambda_\rho \rho)
\end{pmatrix} +
d \begin{pmatrix}
k_z P_z I_{n-1/2}(\lambda_\rho \rho) \\
-\lambda_\rho P_\perp I_{n+1/2}(\lambda_\rho \rho) \\
\Gamma_\lambda I_{n-1/2}(\lambda_\rho \rho)  \\
0
\end{pmatrix}~, \label{wirefuncs}
\end{eqnarray}
where
\bea
\Gamma_k & = & \frac{m_0}{\hbar}\left(E - \frac{E_g}{2}\right) - \frac{\hbar \, m_0}{2 m_\perp^- m_z^-} (k_z^2 m_\perp^- + k_\rho^2 m_z^-)~, \nonumber \\
\Gamma_\lambda & = & \frac{m_0}{\hbar}\left(E - \frac{E_g}{2}\right) - \frac{\hbar \, m_0}{2 m_\perp^- m_z^-} (k_z^2 m_\perp^- - \lambda_\rho^2 m_z^-)~,
\eea
and $a$,$b$,$c$, and $d$ are determined by the boundary conditions.

For NWs with an impenetrable surface, the \textit{standard} boundary conditions require each component of the wave function defined in Eq. \eqref{wirefuncs} to vanish, leading to $\mathcal{R}_i^n(R,k_z)=0$, where $i=1,2,3,4$ and $R$ is the NW radius. These four equations define the 4x4 system for the $a,b,c,d$ coefficients. Requiring the determinant of this system to be zero yields the following implicit equation for the 1D energy bands for angular momentum $n$, and as a function of the parameter $k_z$:
\be
k_\rho  \lambda_\rho \left[(I_+^n)^2 (J_-^n)^2 - (I_-^n)^2 (J_+^n)^2 \right]   +
  \frac{k_z^2 P_z^2 (\Gamma_k - \Gamma_\lambda)^2 +
    P_t^2 (k_\rho^2 \Gamma_\lambda^2-\lambda_\rho^2 \Gamma_k^2)}{P_t^2 \Gamma_k \Gamma_\lambda} I_-^n I_+^n J_-^n J_+^n=0~, \label{det}
\ee
where we use the notation $J_\pm^n = J_{n \pm 1/2}(k_\rho R)$ and $I_\pm^n = I_{n \pm 1/2}(\lambda_\rho R)$.

After determining the energy from Eq. \eqref{det}, the wavefunctions can be constructed from Eq. \eqref{wirefuncs}, with only the normalization undetermined. We will use the following notation for normalized eigenfunctions:  $\Psi_e^{n,k}$ and $\Psi_h^{n,k}$ for the electron and hole levels given by Eq. \eqref{det}, correspondingly, where $k=1,2,3...$ is the index of the 1D subband with angular momentum $n$, and $\int_0^R |\Psi_e^{n,k}|^2 \rho_ed\rho_e 2 \pi = \int_0^R |\Psi_h^{n,k}|^2 \rho_h d\rho_h 2 \pi = 1$.

\begin{figure}[htbp]
    \centering
    \includegraphics[width=8cm]{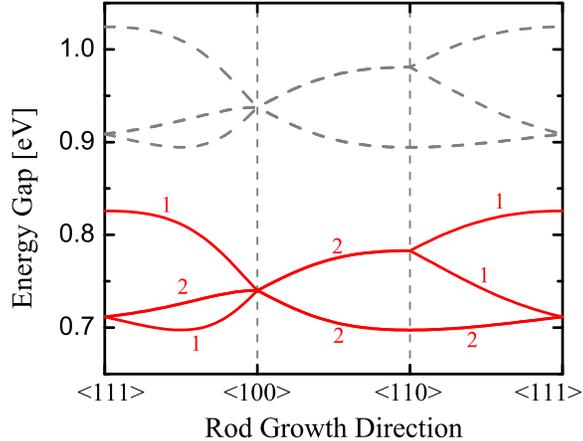}
    \caption{Energy gaps of a 4-nm diameter PbSe NW at each of the four valleys as a function of the growth direction of the NW (red lines). The numbers indicate the valley degeneracy of the energy gaps. Dashed grey lines are the same energy gaps after accounting for the self-Coulomb interaction, described later in the text.}
    \label{enerdirfig}
\end{figure}

Using Eq. \eqref{det} we calculated the energy levels for a 4-nm PbSe NW with various growth directions. The energy band parameters of PbSe which we used in this calculation will be described in a later section. The effective energy gap of the NW, which is the energy distance between the top of the highest 1D sub-band of the valence band and the bottom of the lowest 1D sub-band of the conduction band, impacts many material properties. Figure \ref{enerdirfig} shows the effective energy gap for all four valleys as a function of the growth direction of the nanowire. Because the plot is calculated along high--symmetry directions in the Brillouin zone, the degeneracy of the four valleys is never completely lifted. Without any intervalley coupling, each of these energy gaps would have separate optical absorption and emission peaks associated with it.

\begin{figure}[!htbp]
  \centering
  \begin{minipage}[t]{7.5cm}
    \begin{center}
      \includegraphics[width=7.5cm]{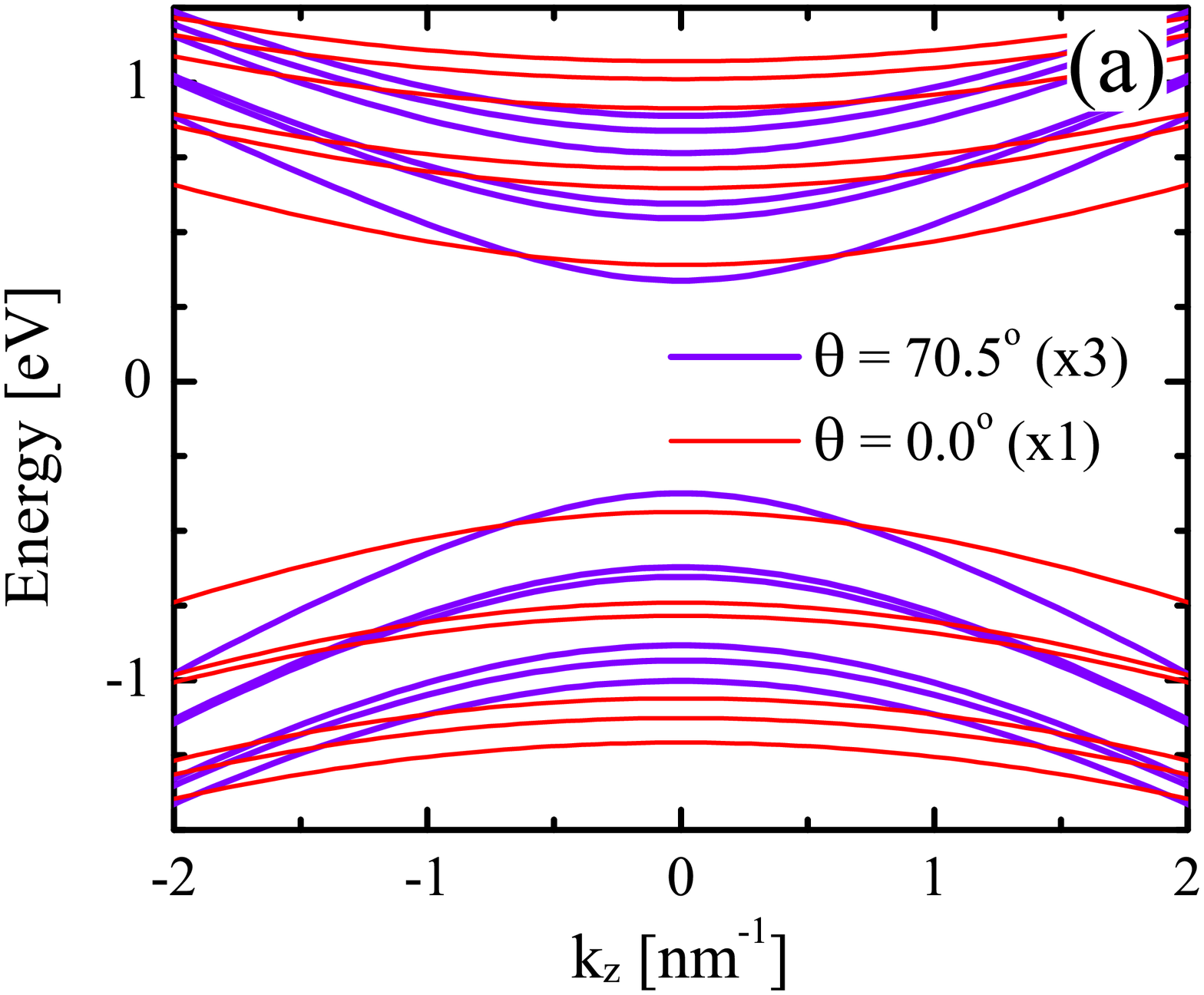}
    \end{center}
  \end{minipage}
  \begin{minipage}[t]{7.5cm}
    \begin{center}
      \includegraphics[width=7.5cm]{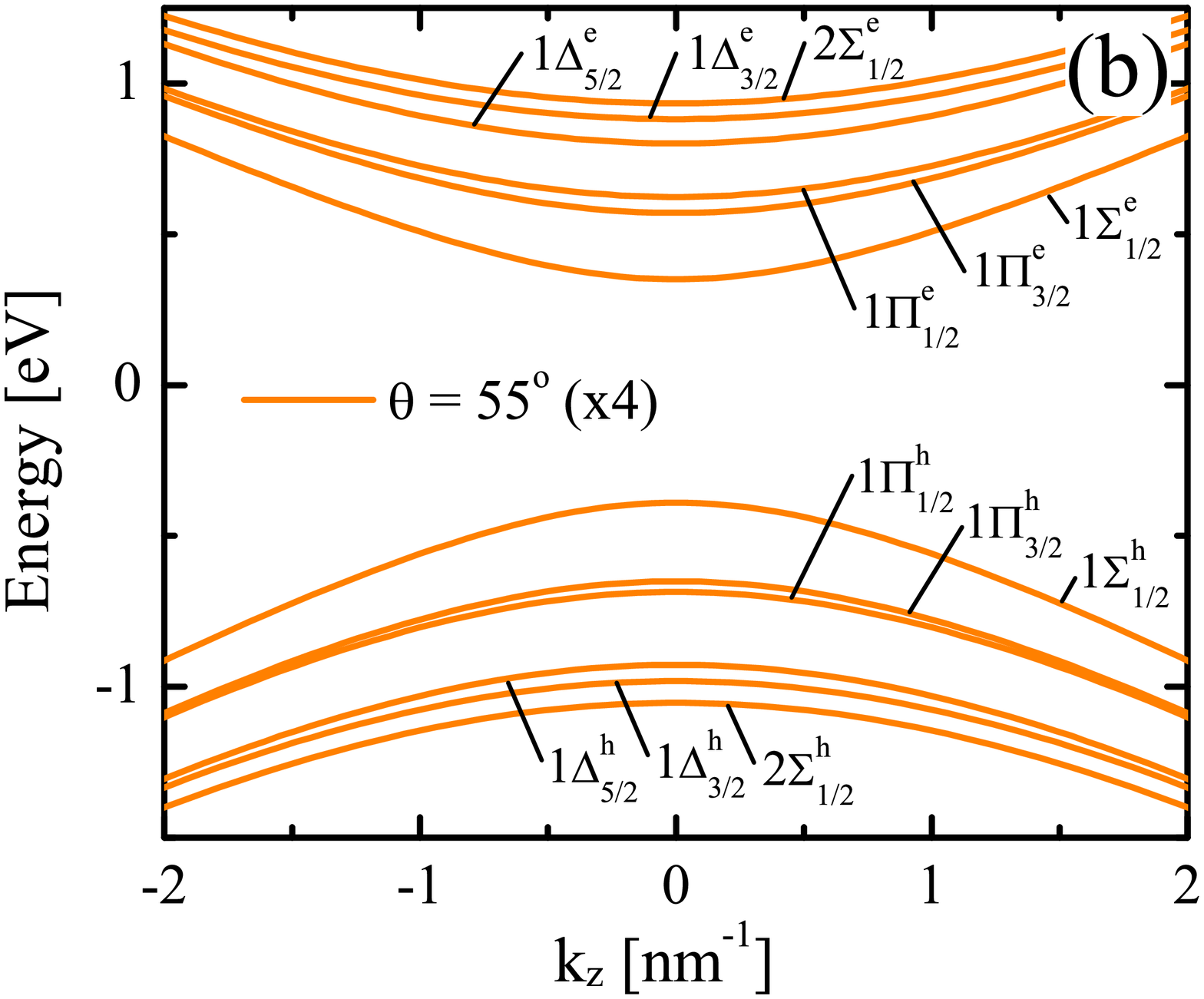}
    \end{center}
  \end{minipage}
  \caption{1D band structure of a 4-nm PbSe NW for the cases of the axis along the directions (a) $\langle 111 \rangle$ and (b) $\langle 100 \rangle$. The bands are labeled by the angle $\theta$ between the considered valley and the rod growth axis and also by their multiple valley degeneracy, up to a maximum of (x4). In (b), the individual subbands are labeled using notation adopted from molecular physics: $k X_{|n|}^{e,h}$ for the $k^{\rm th}$ electron or hole level of certain symmetry with  total $z$ angular momentum $n$, where $X = \Sigma$, $\Pi$, $\Delta$,\ldots, is used for $|m|=0$, 1, 2,\ldots, respectively, where $m$ is the angular momentum projection of the conduction (valence) band component of the  wavefunction of the  electron, `$e$', (hole, `$h$') state. In (a), the order of the levels is the same, and the labeling is suppressed for clarity.}
  \label{bandsfig}
\end{figure}

Figures \ref{bandsfig}a and \ref{bandsfig}b show the dispersion of the several lowest 1D subbands of the conduction and valence bands in NWs grown along the $\langle 111 \rangle$ and $\langle 100 \rangle$ directions,  respectively. NWs grown along $\langle 111 \rangle$ have one valley oriented parallel to the growth direction and the other three valleys oriented at the equal angles $\theta = 71^o$ from it. For the $\langle 100 \rangle$ NW, all four valleys are at the same angle $\theta = 55^o$ from the growth direction. It is clear that both the band-edge energies and the effective masses of the 1D subbands depend strongly on the growth direction.

\section{\label{coulomb} Dielectric confinement}

The optical properties of all semiconductor nanostructures are controlled by the strength of the Coulomb interaction between the electron--hole pair participating in the emission and absorption of photons \cite{EfrosSPS1982}. Compared to the screened Coulomb interaction in a bulk crystal, the interaction is usually enhanced because the electric field of the electron and hole localized inside the nanostructure penetrates into the surrounding medium, which commonly has a dielectric constant smaller than that of the semiconductor. In addition, any charge in the vicinity of this interface polarizes it. In the case of a flat interface, for example, this polarization can be described easily using an image charge that interacts with the primary charge \cite{Jackson}. In the case of small external dielectric constant the interaction is repulsive. This repulsive potential in nanostructures of any shape leads to an additional confinement of carriers, which is referred to as dielectric confinement.

To model these effects in NRs and NWs, the analytic potential for two charges in an infinite dielectric cylinder $U({\bm r}_e,{\bm r}_h)$ \cite{smythe} is used. It was shown previously \cite{shabaev_efros} that this approximation works well as long as the rod length is larger than the size of the exciton. The potential naturally divides into four terms \cite{brus}: the unscreened direct interaction of the two charges $U_d$, the modification of this interaction due to the image effects of the solvent $U_s$, and the two self-interactions of each charge with its own image $U_e$ and $U_h$:
\begin{equation}
\begin{matrix}
U({\bm r}_e,{\bm r}_h) & = & -e^2/(\kappa_s | {\bm r}_e - {\bm r}_h|) & - & e V_s({\bm r}_e,{\bm r}_h) & + & e V_s({\bm r}_e,{\bm r}_e)/2 & + & e V_s({\bm r}_h,{\bm r}_h)/2\label{bindingeq}\\
& \equiv & U_d (| {\bm r}_e - {\bm r}_h|)& + & U_s({\bm r}_e,{\bm r}_h) & + & U_e({\bm r}_e) & + & U_h({\bm r}_h) \end{matrix}
\end{equation}
where the function $V_s$ has the form
\begin{eqnarray}
V_s({\bm r}_e,& &{\bm r}_h) = \frac{e}{2 \pi^2 \kappa_s}  \int_{0}^\infty \, du \sum_{m=0}^\infty \cos(u (z_e - z_h)) \cos(m (\phi_e - \phi_h)) (2 - \delta_{m0})\times \label{Vs}\\
& & \times \frac{(\kappa_s-\kappa_m)I_m(u \rho_e) I_m(u \rho_h)K_m(R u)\left(K_{m-1}(R u) + K_{m+1}(R u)\right)}{\kappa_s K_m(R u) \left(I_{m-1} (R u) + I_{m+1} (R u)\right) + \kappa_m I_m(R u) \left(K_{m-1}(R u) + K_{m+1}(R u) \right)}\nonumber
\end{eqnarray}
and where $\kappa_s$ and $\kappa_m$ are the optical dielectric constants of the bulk semiconductor and the surrounding medium, respectively.  $I_m$ and $K_m$ are the modified Bessel functions of the first and second kind. For PbSe we will use $\kappa_s = 23$, and for the medium, if not explicitly stated otherwise, $\kappa_m = 2$ throughout this work.

The self-interaction terms  $U_e({\bm r}_e)$ and $U_h({\bm r}_h)$ always contribute to the energy of each electron and hole subband calculated in Section \ref{nanowire}. In narrow NWs and NRs, where the self--interaction energy is smaller than the confined energies, this contribution can be calculated perturbatively for electron and hole levels, respectively:
\be
E_\text{self,e}^{n,k}  =  \int \rho_e d\rho_e d\phi_e |\Psi_e^{n,k}|^2U_e({\bm r}_e)~,~  E_\text{self,h}^{n',k'}  =\int \rho_h d\rho_h d\phi_h  |\Psi_h^{n',k'}|^2 U_h({\bm r}_h)\label{Eself}~.
\ee
The self-interaction terms $E_\text{self,e}^{n,k}$ and $E_\text{self,h}^{n',k'}$ increase the energy of all electron and hole 1D subbands and consequently the effective energy gap in nanowires. The perturbed electron and hole subbands with $n=n'=1/2$ and $k=k'=1$ are shown in Fig. \ref{enerdirfig}.

\begin{figure}[htbp]
    \centering
    \includegraphics[width=8cm]{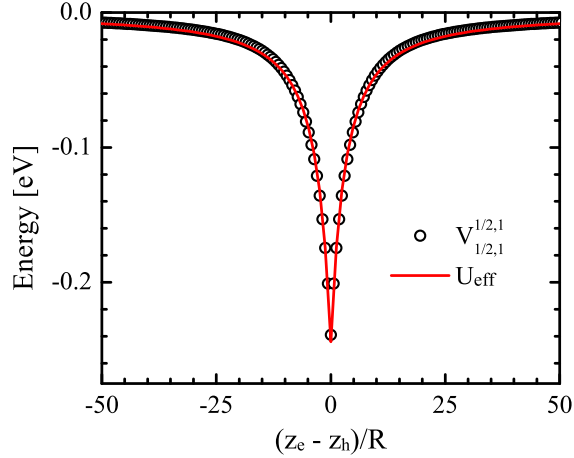}
    \caption{Points show the effective binding potential, $V_{1/2,1}^{1/2,1}$, between an electron and a hole occupying the ground one dimensional subband $n=n'=1/2$ and $k=k'=1$ as a function of their separation, calculated for a 4-nm radius PbSe NW. The solid line shows the approximation of this dependence by the Elliott \& Loudon effective potential described by Eq. \eqref{Ueff}}
    \label{Veff}
\end{figure}

In addition, in narrow NWs and NRs one can used an adiabatic approximation of the Coulomb interaction \cite{loudon,elliott_loudon}, which replaces the three-dimensional potential of electrons and holes of Eq. \eqref{bindingeq} by a one-dimensional Coulomb potential that describes their interaction along the NW/NR axis. The adiabatic  potential is obtained by averaging the potential over wave functions $\Psi_e^{n,k}$ and $\Psi_h^{n',k'}$ of the corresponding electron and hole subband. Averaging the first two terms of Eq. \eqref{bindingeq} results in the 1D adiabatic potential
\be
V_{n,k}^{n'k'}(|z_e - z_h|)  = \int \rho_e d\rho_e d\phi_e \int \rho_h d\rho_h d\phi_h  |\Psi_e^{n,k}|^2 |\Psi_h^{n',k'}|^2 (U_d(| {\bm r}_e - {\bm r}_h|) +  U_s({\bm r}_e,{\bm r}_h))~, \label{Vavg}\\
\ee
which describes the interaction of electrons and holes occupying different subbands. This adiabatic potential is a function of the electron and hole separation,  $|z_e - z_h|$, only.  One can show that at large distances $|z_e - z_h|\gg R$ it takes the form of a one-dimensional Coulomb potential with the dielectric constant of the surrounding medium, $V_{n,k}^{n'k'} \sim -e^2/(\kappa_m|z_e - z_h|)$. The adiabatic potential for the ground electron and hole subbands with $n=n'=1/2$ and $k=k'=1$ is shown in Fig.\ref{Veff}.

\section{\label{1Dexc}1D excitons in PbSe nanowires and nanorods}

The attractive 1D potential described by Eq. \eqref{Vavg} creates a series of one-dimensional exciton states for each pair of electron and hole subbands $(n,k)$ and $(n',k')$.  The effective masses of electrons and holes along the NW axis $m_e^{n,k}$ and $m_e^{n',k'}$ at the bottom and the top of each subband, correspondingly, is determined by Eq. \eqref{det}. This allows us to write a one-dimensional  Schr\"{o}dinger equation for these 1D excitons:
\be
-\frac{\hbar^2}{2 \mu_{n,k}^{n'k'}} \frac{\partial^2}{\partial z^2} \Psi_\text{1D} -\frac{\hbar^2}{2 M_{n,k}^{n'k'}} \frac{\partial^2}{\partial Z^2} \Psi_\text{1D} +  U_{n,k}^{n'k'}(z)  \Psi_\text{1D}= \varepsilon_{n,k}^{n'k'} ~ \Psi_\text{1D}~,
\label{SE}
\ee
where we introduce the electron-hole separation, $z = z_e - z_h$ and the exciton center-of-mass coordinate $Z = (m_e^{n,k} z_e + m_h^{n',k'} z_h) / (m_e^{n,k} + m_h^{n',k'})$. $\mu_{n,k}^{n'k'} = m_e^{n,k} m_h^{n',k'} / (m_e^{n,k} + m_h^{n',k'})$ is the reduced mass and $M_{n,k}^{n'k'} = m_e^{n,k}  + m_h^{n',k'}$ is the total effective mass of the 1D exciton. Importantly, the exciton binding energy $\varepsilon_{n,k}^{n'k'}$ in this equation is calculated relative to the distance between the bottom of the $(n,k)$ conduction subband and the top of the $(n',k')$ valence subband, assuming the self--interaction energy terms  $E_\text{self,e}^{n,k}$ and $E_\text{self,h}^{n',k'}$ are already taken into account. The solution of Eq. \eqref{SE} can be separated into $\Psi_\text{1D}(z,Z) = \psi_\text{1D}(z)\Psi_\text{cm}(Z)$. The wave function $\psi_\text{1D}(z)$ describes relative electron-hole motion and gives the spectrum of 1D excitons. The second component, $\Psi_\text{cm}(Z)$, describes the exciton center of mass motion, and in the case of an infinite NW $\Psi_\text{cm}(Z)\sim \exp(iKZ)$, where $\hbar K$ is the exciton momentum along the NW axis.  This replaces the second term in Eq. \eqref{SE} by the exciton kinetic energy, $\hbar^2K^2/2 M_{n,k}^{n'k'}$.

\begin{figure}[!htbp]
  \centering
  \begin{minipage}[t]{7.5cm}
    \begin{center}
      \includegraphics[width=7.5cm]{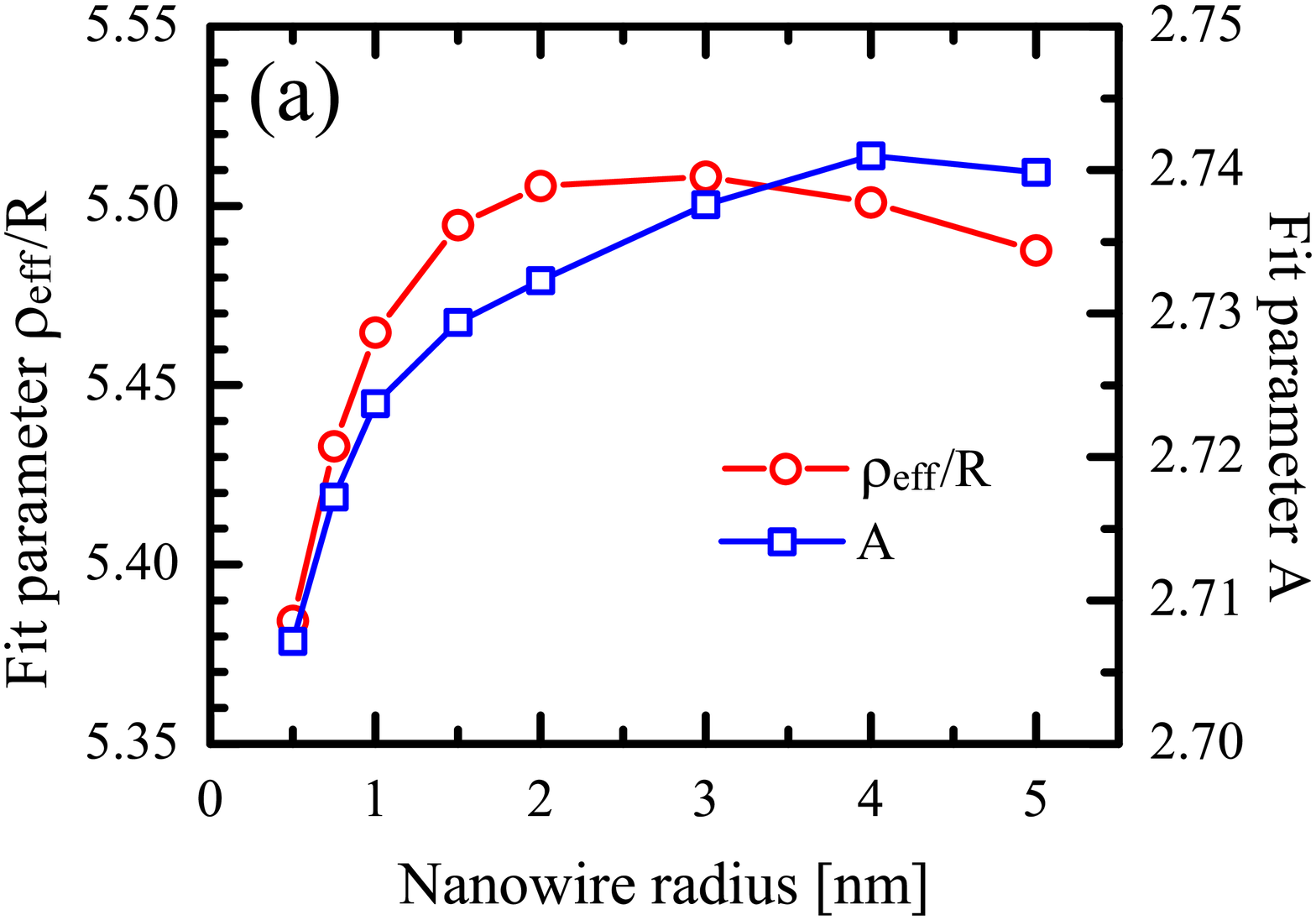}
    \end{center}
  \end{minipage}
  \begin{minipage}[t]{7.5cm}
    \begin{center}
      \includegraphics[width=7.5cm]{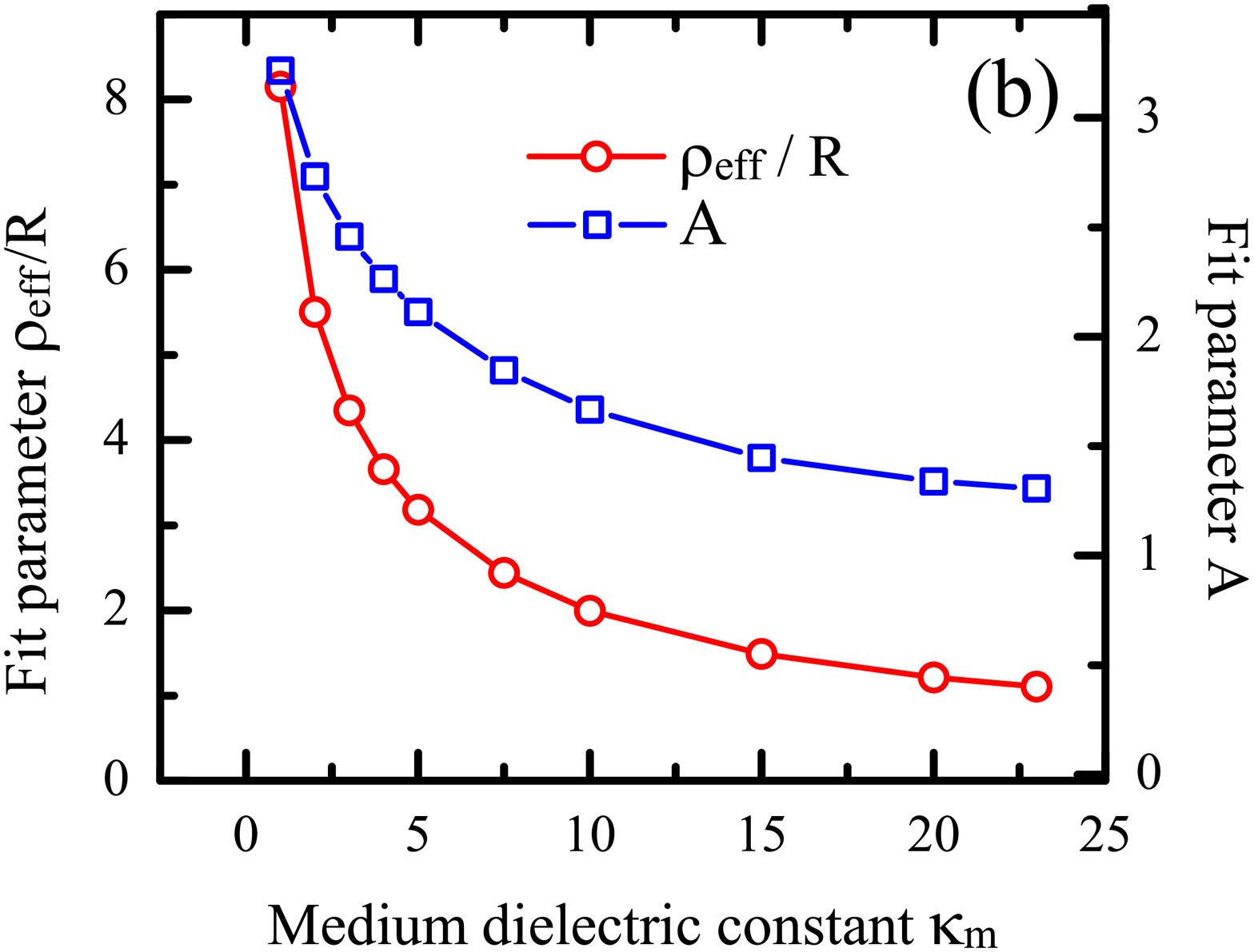}
    \end{center}
  \end{minipage}
  \caption{Fitting parameters used in the effective potential described by Eq. \eqref{Ueff} in PbSe NWs of various radius and medium dielectric constant. The parameter is plotted vs. (a) nanowire diameter with $\kappa_m = 2$ (b) medium dielectric constant with $R = 2$ nm.}
  \label{potparams}
\end{figure}

Equation \eqref{SE} allows us to numerically calculate the energy spectrum of 1D excitons created from any pair of electron and hole subbands.  In this paper, we will be interested primarily in the spectrum that arises from the lowest electron and hole subbands $1\Sigma^e_{1/2}$ and $1\Sigma^h_{1/2}$, and we will use the approach suggested by Elliott \& Loudon \cite{elliott_loudon} to describe the spectrum of one-dimensional excitons in a strong magnetic field. They suggest approximation of the one-dimensional adiabatic potential by an effective one-dimensional potential, which has well-known Schr\"{o}dinger equation solutions,
\begin{equation}
U_\text{eff}(z) = - \frac{e^2}{\kappa_m ( |z| + \rho_\text{eff} )}- \frac{A \rho_\text{eff} \, e^2}{\kappa_m ( |z| + \rho_\text{eff} )^2}~, \label{Ueff}
\end{equation}
where $\rho_\text{eff}$ and $A$ are the two fitting parameters. The medium dielectric constant $\kappa_m$ is used in Eq. \eqref{Ueff} so that the correct asymptotic form of the potential is maintained. For a 4-nm PbSe NW immersed in a medium with $\kappa_m = 2$, the numerically-calculated effective potential is described very well by the potential $U_\text{eff}$ with $\rho_\text{eff}=5.49 R$ and $A=2.73$, as seen in Fig. \ref{Veff}. The slight dependence of these fit parameters on NW size is shown in Fig. \ref{potparams}a and the much stronger dependence on $\kappa_m$ is shown in Fig. \ref{potparams}b.

The energy spectrum and eigenfunctions of Eq.\eqref{SE} with effective attractive potential $U_\text{eff}(z)$ can be obtained analytically. The eigenfunctions of each 1D exciton level, $\psi_\alpha(z)$, can be written as \cite{loudon, elliott_loudon}
\bea
\psi_\alpha(z > 0) & = & a_1 W_{\alpha, -\frac{1}{2}\sqrt{1 - 4 A \alpha \tilde{\rho}}}(\tilde{z}+\tilde{\rho})+a_2M_{\alpha, -\frac{1}{2}\sqrt{1 - 4 A \alpha \tilde{\rho}}}(\tilde{z}+\tilde{\rho})~ \label{WF} \\
\psi_\alpha(z < 0) & = & \pm \psi_\alpha(|z|) \label{psizlt0}
\eea
where $W_{\alpha,\beta}(x)$ and $M_{\alpha,\beta}(x)$ are the Whittaker functions, $\tilde{z} =2 z/(a_0 \alpha)$, $\tilde{\rho} = 2 \rho_\text{eff}/(a_0 \alpha)$, $a_0=\hbar^2\kappa_m/(\mu_{1/2,1}^{1/2,1} e^2)$  is the effective Bohr radius of a 1D exciton, and $a_1$ and $a_2$ are arbitrary coefficients. The sign of Eq. \eqref{psizlt0} is ``$+$'' for an even eigenfunction and ``$-$'' for an odd one. The coefficients $a_1$, $a_2$, and parameter $\alpha$ in Eq. \eqref{WF} as well as the exciton binding energy:
\be
\varepsilon_\alpha = - \frac{\hbar^2}{2 \mu_{1/2,1}^{1/2,1} a_0^2 \alpha^2}
\label{BE}
\ee
are determined by the  boundary conditions.

There are two boundary conditions to impose on the solution in Eq. \eqref{WF}: one at $z = z_e - z_h = \pm L$ and one at $z = 0$. We first consider infinite nanowires; the effects of finite length will be treated in the following section. In this case, the first boundary condition is satisfied by letting $a_2=0$, because $M_{\alpha, -\frac{1}{2}\sqrt{1 - 4 A \alpha \tilde{\rho}}}(|\tilde{z}|+\tilde{\rho})$ diverges as $\tilde{|z|}\rightarrow \infty$. The second boundary condition, requiring $\psi_\alpha(z)$ to be either an even or odd function of $z$, determines $\alpha$ and the energy spectrum of the exciton. It was shown in Refs. \inlinecite{loudon, elliott_loudon} that for excited doubly-degenerate exciton states, $\alpha$ takes almost-exactly integer values $\alpha=1,2,3,...$ and that $\alpha \rightarrow 0$ for ground states with decreasing exciton transverse radius. Following  Refs. \inlinecite{loudon, elliott_loudon} we use $\varepsilon_0$ for the ground exciton binding energy.

\begin{figure}[!htbp]
  \centering
  \begin{minipage}[t]{7.5cm}
    \begin{center}
      \includegraphics[width=7.5cm]{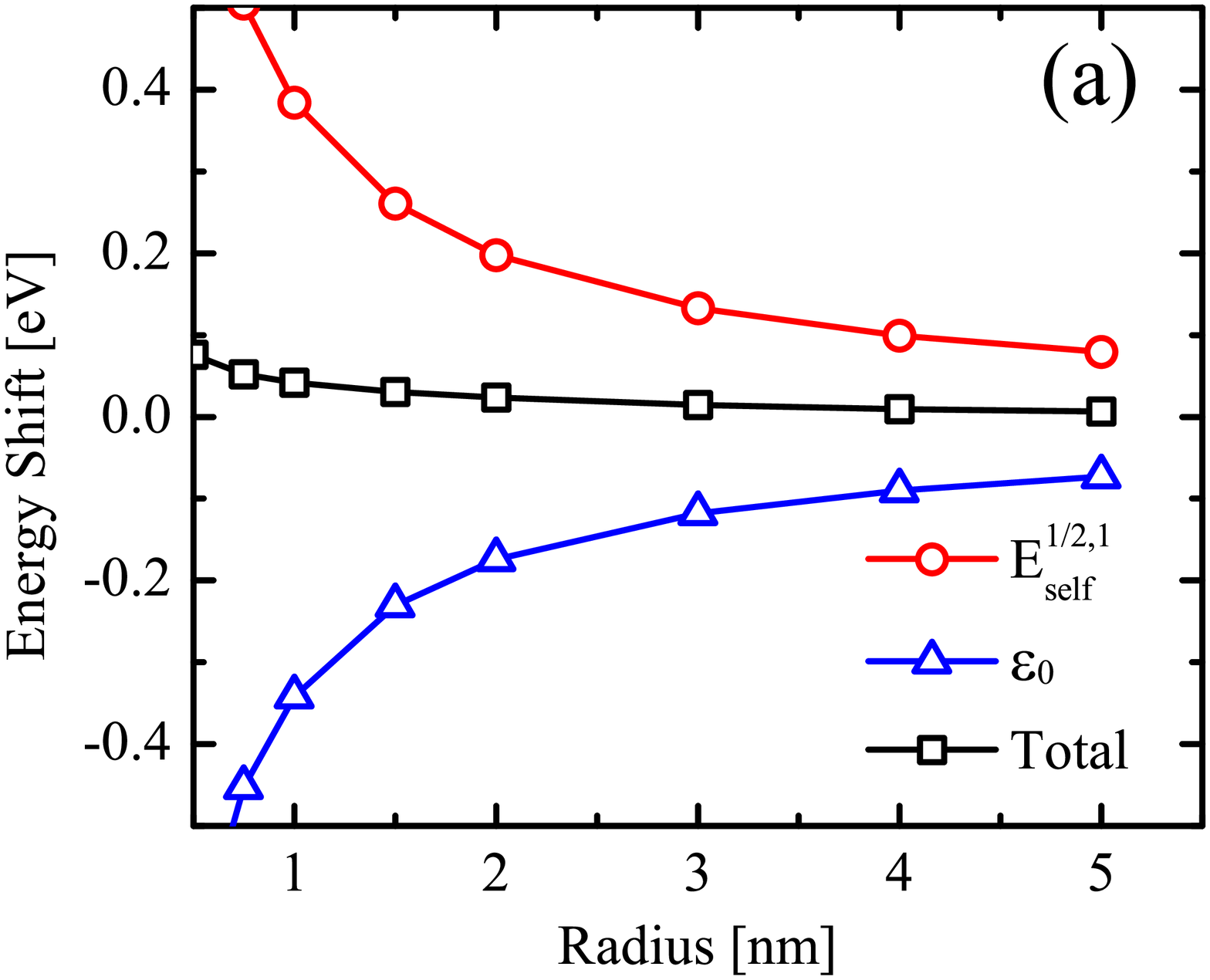}
    \end{center}
  \end{minipage}
  \begin{minipage}[t]{7.5cm}
    \begin{center}
      \includegraphics[width=7.5cm]{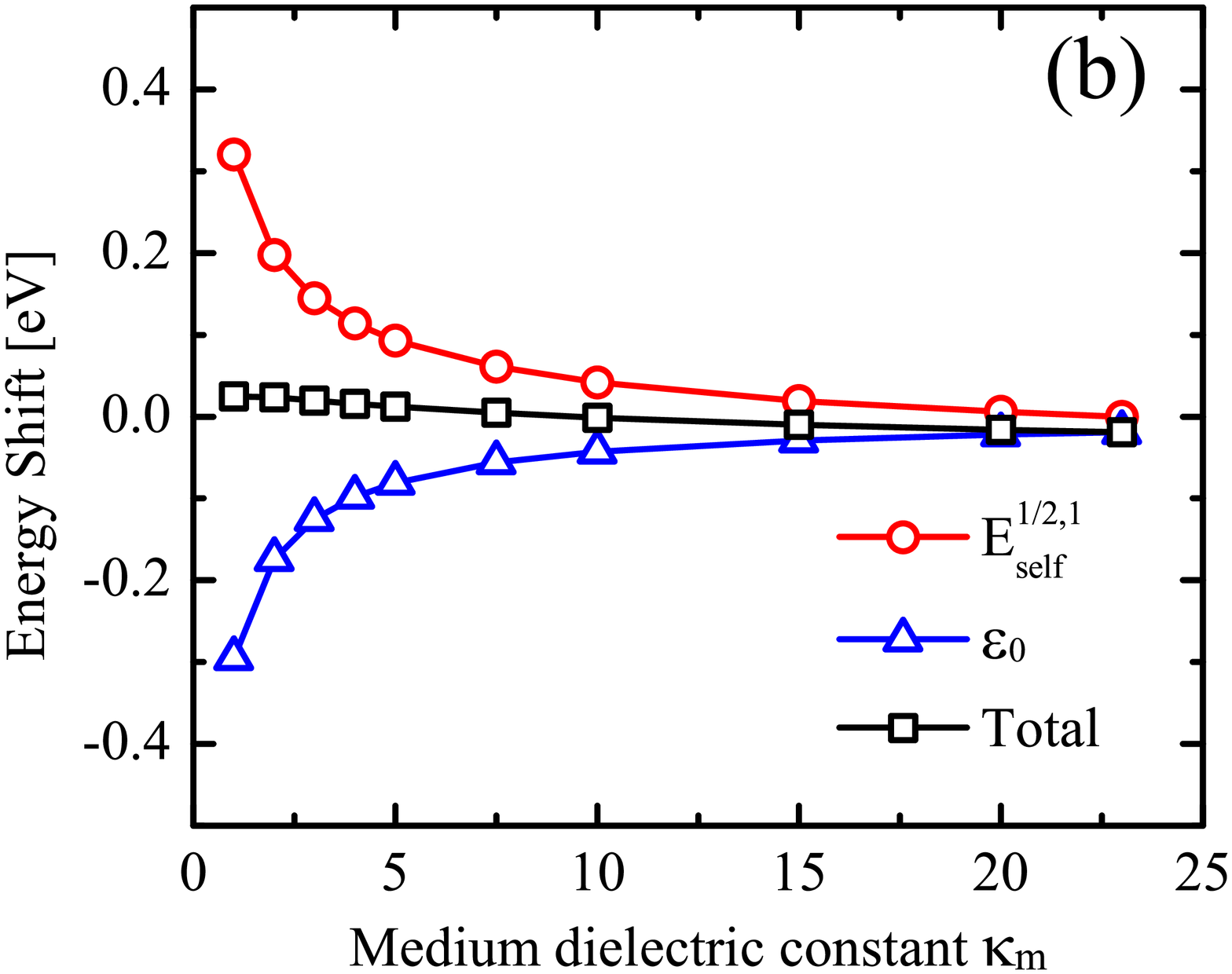}
    \end{center}
  \end{minipage}
  \caption{Coulomb energies calculated for (a) $\kappa_m = 2$ with varying $R$ and (b) $R = 2$ nm with varying $\kappa_m$. Lines are the sum of the electron $E_\text{self,e}^{1/2,1}$ and hole $E_\text{self,h}^{1/2,1}$  self interaction energies (red circles); the electron-hole binding energy $\varepsilon_0$ (blue triangles); and their total (black squares).}
  \label{bindingener}
\end{figure}

Figure \ref{bindingener} shows the calculated binding energy of the ground exciton state $\varepsilon_0 $ and the Coulomb self--interaction energies $E_{{\rm self},e}^{1/2,1}$ and $E_{{\rm self},h}^{1/2,1}$ of electrons and holes from the ground 1D subbands $1\Sigma^{e,h}_{1/2}$. The binding energy decreases dramatically with NW radius or external dielectric constant.  The exciton binding energy in the narrowest NW surrounded with $\kappa_m\sim 2-3$ reaches values $>300$\,meV.

Surprisingly, however, the binding energy is almost exactly compensated by the electron and hole self--interaction terms, which leads to practical cancelation of most effects connected with the small dielectric constant of the surrounding medium. Because of this cancelation, the optical transitions between 1D subbands will be determined primarily by the energies calculated in Section \ref{nanowire}. This result has important practical consequences.  For example, the linear optical spectra of PbSe NWs will not be sensitive to the dielectric constant of the surrounding medium.

This cancelation is well--known in spherical semiconductor NCs. The exact cancelation of these three terms was shown for parabolic valence and conduction bands in Ref. \inlinecite{ChemlaSRingMillerPRB88}.  This is because in a parabolic-band approximation the wave function of electrons and holes are identical and depend only on the NC radius. As a result the electron and hole charge distributions exactly compensate each other at each point in the NC. If there is no local charge in the NC, there is no electric field outside of the NC, and the external medium does not affect the optical properties. This cancelation is nearly exact even when the electron and hole masses are different \cite{efrosSSC88}.

\begin{figure}[htbp]
    \centering
    \includegraphics[width=8cm]{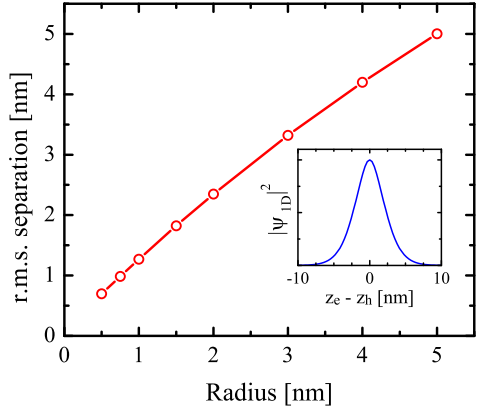}
    \caption{Dependence on PbSe NW radius of the average (r.m.s.) separation of the electron and hole in the 1D exciton. Inset shows the square of  the ground exciton wavefunction $|\psi_{1D}|^2$  for a NW with 2 nm radius.}
    \label{separation}
\end{figure}

The cancelation of the Coulomb energies in the ground exciton of PbSE NWs can be attributed to a similar charge compensation. The mirror symmetry of the conduction and valence bands in PbSe makes the wave functions of the electron and hole transverse motion nearly identical.  The similar values of effective masses along the NW axes also makes the electron and hole contributions to the 1D exciton wave function identical. It is interesting to note here that because of the large binding energy, the electron and hole in the exciton are remarkably tightly bound, with average separation only slightly larger than the NW radius. Fig. \ref{separation} shows the average separation, calculated as $\sqrt{\langle (z - \bar{z})^2 \rangle}$,  as a function of radius, with inset showing the wavefunction $\psi_{1D}$ for the case of $R = 2$ nm. One can see that the average electron-hole separation in the exciton is an order of magnitude smaller than the 46 nm Bohr radius in bulk PbSe. Further calculations show that this unusual increase in the strength of the binding is due entirely to the 1D shape of the NR, and is only weakly affected by the dielectric contrast.  For the weakest dielectric contrast when $\kappa_m = \kappa_s = 23$, the average separation increases slightly to $\approx 4$ nm, still much closer to the 4-nm diameter than to the Bohr radius.

\subsection{Finite length effects \label{finitelength}}

For a nanorod, which has finite length, the relative and center-of-mass (CM) motions of the electron and hole can never be completely separated. If the NR is much longer than the radius of the 1D  exciton, one  can  still  approximately separate variables to create effective boundary conditions for the exciton CM motion. No other boundary condition (BC) is needed for the exciton separation coordinate, because the assumption is that the tightly--bound wavefunction is already zero well before any additional confinement is felt. On the other hand, the CM motion can be considered as the motion of a free particle confined in a 1D box of length $L$.  If the box is much larger than the exciton radius one can apply the standard boundary conditions on $\Psi_\text{cm}$ to obtain the well-known spectrum $E_\text{cm}(l)=\hbar^2\pi^2l^2/(2M_{1/2,1}^{1/2,1}L^2)$, where $l$ is the level number.

\begin{figure}[htbp]
    \centering
    \includegraphics[width=8cm]{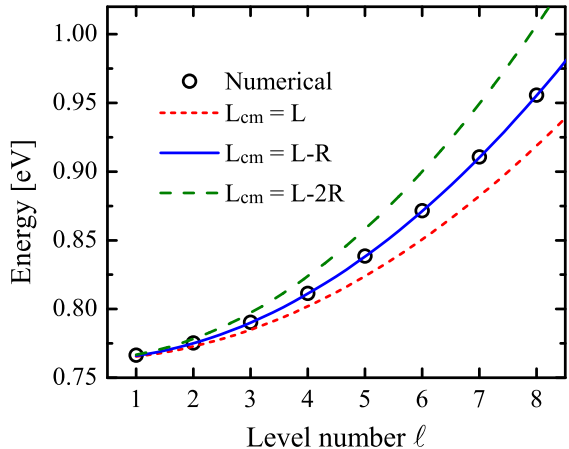}
    \caption{Numerically-calculated energies for the lowest few exciton states in a 4 x 20 nm PbSe NR (black circles). The lines are the energies from the analytic model using two different confinement lengths for the center of mass.}
    \label{COMenergies}
\end{figure}

Even though this CM boundary condition makes intuitive sense, it is difficult to justify, because the true BCs are for the electron and hole individually. To test our assumption, we calculated the CM energies numerically by solving the two-particle Schr\"{o}dinger equation with the correct impenetrable boundary conditions on the electron and hole individually. Details of the calculation are in Appendix \ref{comappendix}. The numerically calculated wavefunctions and energies were best matched to those obtained for a free particle with an effective mass of the exciton  which is  confined in the 1D box of length $L_\text{cm} = L- R$. The existence of such a simple expression is probably connected with the approximately-equal effective masses of the electrons and holes and their small separation in PbSe NRs. The first few numerically-calculated energy levels are shown in Fig. \ref{COMenergies}, along with the analytic energies $E_\text{cm} = \hbar^2 l^2 \pi^2/(2 M_{1/2,1}^{1/2,1} L_\text{cm}^2)$ for various confinement lengths $L_\text{cm}$. This modified CM length works well for all rod sizes studied, as long as the NR aspect ratio is $\gtrsim2$.

\subsection{\label{OTR}Oscillator strength of the interband optical transitions}

The decrease of the electron-hole separation within a 1D exciton leads to a dramatic increase of the optical transition strength. It was shown by Elliott \& Loudon \cite{elliott_loudon} that  the oscillator strength of practically the entire spectrum of 1D excitons becomes concentrated in the ground exciton state. The expression for the transition strength in PbSe NRs can be obtained by combining the results derived for PbSe NCs \cite{kang_wise} and CdSe NRs \cite{shabaev_efros}. The total oscillator strength $O_\text{total}$ can be written as a product $O_\text{total} = O_\perp O_\parallel$, where the tranverse oscillator strength is \cite{kang_wise}
\be
O_\perp  =  \frac{2 P_l^2}{9 m_0 \hbar\omega} \left| \int_0^R \rho d\rho \int_0^{2 \pi} d \phi
\Bigg[ \Psi_h^{1/2,1} \Bigg]^\dagger
\Bigg[\begin{matrix}
0 & \sigma_z \\
\sigma_z & 0
\end{matrix} \Bigg]
\Bigg[\Psi_e^{1/2,1} \Bigg] \right|^2 \label{oscwire}
\ee
with $\hbar\omega$ the total energy of the optical transition. We have neglected the second term from Ref. \inlinecite{kang_wise}, as it is negligible except for very small NRs, where the envelope function approximation likely breaks down anyway. The oscillator strength of the 1D exciton \cite{shabaev_efros} is
\be
O_\parallel =   |\psi_{1D}(z=0)|^2\left| \int_0^L \, dZ  \Psi_\text{cm}(Z) \right|^2 \label{oscrod}
\ee
where we normalize the 1D exciton wave function such that $\int_{-L}^L \, d z \int_0^L \, d Z |\psi_{1D}(z) \Psi_\text{cm}(Z)|^2=1$.

The transverse oscillator strength provides the selection rule that there is no change in the z-component of the angular momentum, $\Delta n = 0$, while the longitudinal component focuses the oscillator strength into the ground exciton state. This is because optical transitions are only allowed to the even states of the exciton CM motion with $l=1,3,5...$, and the oscillator strength decreases as $1/l^2$. Even the second allowed transition will be 9 times weaker than the lowest transition. This has practical implications for the optical absorption spectra. Even though the density of allowed transitions increases dramatically with energy in NRs, most of the oscillator strength is concentrated in the lowest-energy transition for each pair of NR subbands. Thus, isolated peaks should still be observable in experimental spectra.

\section{\label{experiment}Experiment}

\subsection{Synthesis and characterization of colloidal PbSe nanorods}

Although the synthesis of lead salt nanowires was reported several years ago \cite{efrat, cho_murray}, the fabrication of high quality lead-salt nanorods with small diameter has proved challenging. PbSe NRs were synthesized with noble metals as seeds \cite{yong_prasad}, but the resulting NRs did not have good optical spectra.  Some high-quality NRs have been reported, but the syntheses were too challenging for us to reproduce \cite{luther, acharya, warner}. A simple synthesis for high-optical-quality PbSe NRs was recently demonstrated \cite{koh}, and the properties of these NRs will be compared to the theoretical results.

Following Ref. \inlinecite{koh}, the NR synthesis was carried out using standard Schlenk-line techniques under dry nitrogen. Tris(diethylamino)phosphine (TDP, Aldrich, 97\%), oleic acid (OA, Aldrich, 90\%), 1-octadecene (ODE, Aldrich, 90\%), squalane (Aldrich, 99\%), amorphous selenium shots (Se, Aldrich, 99.999\%), and lead(II) oxide (PbO, Aldrich, 99.9\%) were used as purchased without further purification. Anhydrous ethanol, chloroform, acetone, hexane, and tetrachloroethylene (TCE) were purchased from various sources. To prepare 1.0 M stock solutions of TDPSe, 7.86 g of Se was dissolved in 100 mL of TDP.

Typically, 0.22 g of PbO was dissolved in 5 mL of squalane in the presence of 1 mL OA. (Squalane can be replaced by ODE.) After drying under nitrogen at 150 °C for 30 min, the solution was heated to 170 °C and 3 mL of a 1 M TDPSe solution in TDP was injected under vigorous stirring. Once the reaction finished, the reaction mixture was cooled to room temperature using a water bath. The crude solution was mixed with hexane and precipitated by ethanol. The precipitated NRs were isolated by centrifugation (at 5000 rpm for 3 min) and redispersed in chloroform or other organic solvents. Size-selective precipitation can be carried out to obtain better monodispersity of NRs samples using chloroform/acetone or other solvent/nonsolvent pairs.

The size of the synthesized NRs was determined from transmission electron microscopy. In-plane powder X-ray diffraction shows that the NRs grow along the $\langle 100 \rangle$ direction \cite{koh}. Absorption was measured on a Shimadzu UV-3101PC spectrophotometer at room temperature. Emission spectra were recorded at room temperature with an infrared fluorimeter equipped with a 200-mm focal length monochromator, a single mode fiber coupled laser source (S1FC635PM, 635 nm, Thorlabs, Inc) as the excitation source, and an InGaAs photodiode (New Focus Femtowatt model 2153). Fluorescence lifetime was measured using an InP/InGaAs PMT (Hamamatsu H10330A-75) with 120-fs excitation pulses from a Ti:sapphire regenerative amplifier (Spectra-Physics Hurricane) with 1 kHz repetition rate. NRs were dissolved in tetrachloroethylene (TCE) for all measurements to avoid spurious absorbance in the near-IR. Quantum yield measurements were performed using an integrating sphere, with the method described in Ref. \inlinecite{qysphere}.

\subsection{Absorption Spectra}

\begin{figure}[!htbp]
  \centering
  \begin{minipage}[t]{7.5cm}
    \begin{center}
      \includegraphics[width=7.5cm]{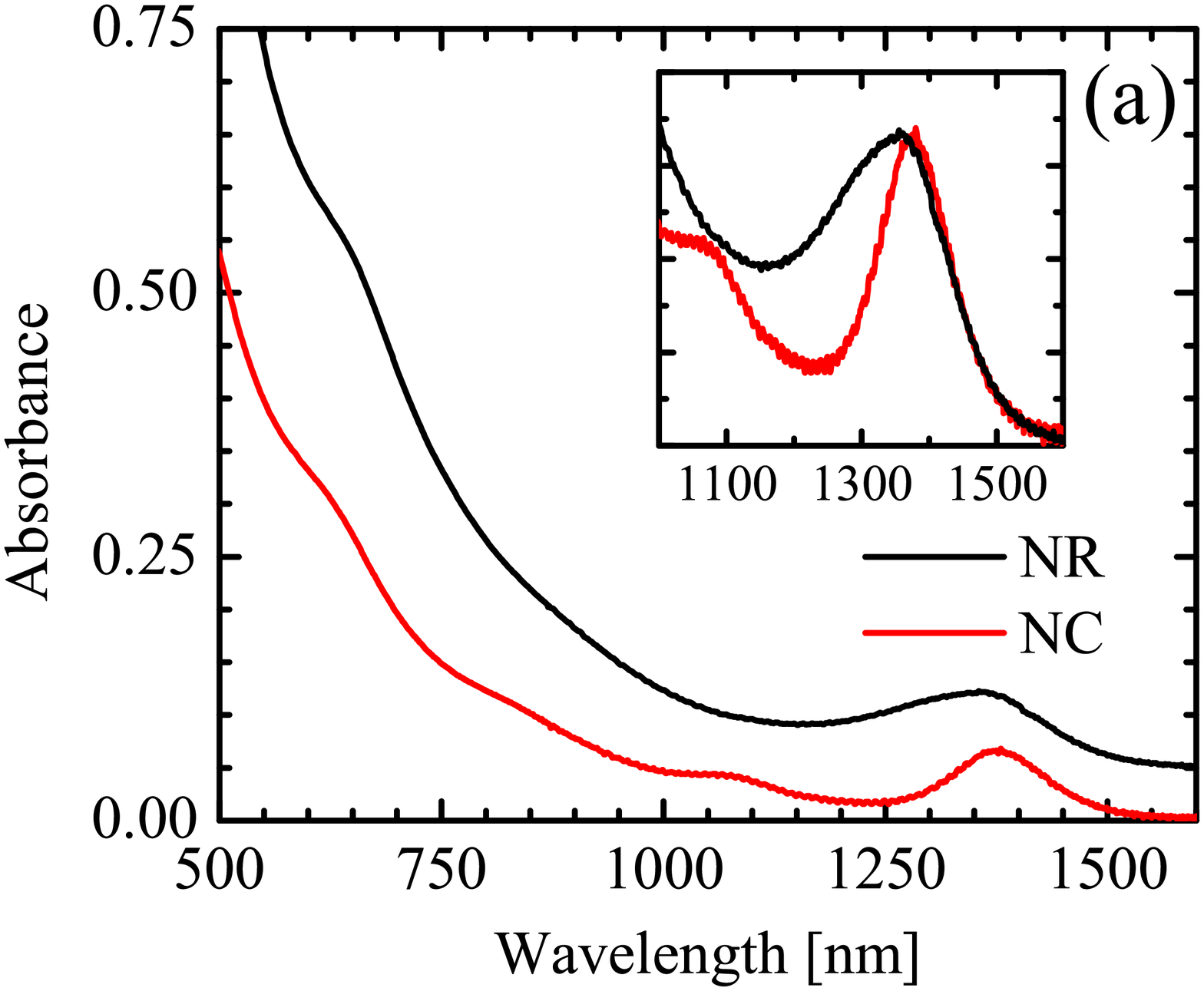}
    \end{center}
  \end{minipage}
  \begin{minipage}[t]{7.5cm}
    \begin{center}
      \includegraphics[width=7.5cm]{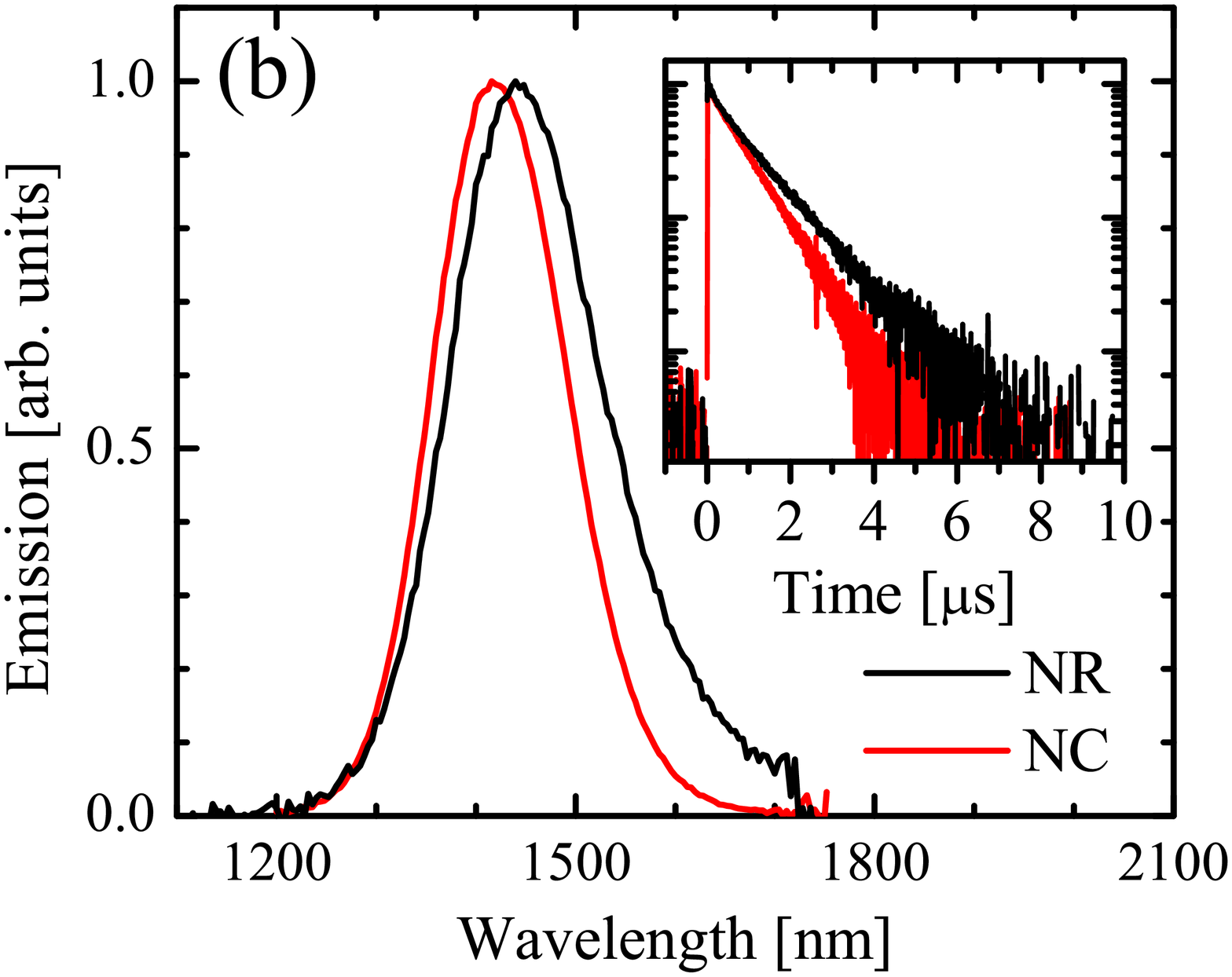}
    \end{center}
  \end{minipage}
  \caption{(a) Absorption spectra of PbSe NRs (black line, vertically offset for clarity) and spherical PbSe NCs (red line) are compared. The inset shows detail of the first peak. (b) Emission spectra and fluorescence decays measured at the emission peak (inset) of the same two samples.}
  \label{QDvsNRgraph}
\end{figure}

First, we will highlight the qualitative differences between the absorption spectra of NRs and spherical NCs. Figure \ref{QDvsNRgraph} shows the absorption spectrum of 3.3 nm diameter x 12 nm length PbSe NRs along with that of 4.4 nm diameter spherical NCs, chosen to have a nearly identical first absorption peak. The spectrum of the NRs has fewer obvious features than the NC spectrum. The first peak in the NR spectrum has a broad high energy side, even though its narrower low energy side is nearly identical to that of the NCs (inset of Fig. \ref{QDvsNRgraph}). Both of these observations indicate the presence of more densely-spaced transitions in the NR spectrum, which have the effect of smoothing out the peaks. Interestingly, the second NC peak appears where there is a dip in the NR spectrum.

The broadening of the NR absorption peak seen in Fig. \ref{QDvsNRgraph} is connected with the dispersion of NR diameter and length.  Our best PbSe NR samples have around 5\% size distribution in radius, but a much larger 20\% in length. This large length polydispersity will blur out many of the NR transitions in an ensemble, except for those that are roughly independent of length--- specifically, the lowest energy exciton for each pair of NW subbands. Fortunately, this is also the transition predicted to have the largest oscillator strength. As we have shown above, the energies of the optical transitions of the ground exciton states practically coincide with the energies between non-interacting electron and hole subbands% due to the dielectric screening effect
, even though their respective wave functions differ greatly. This greatly simplifies the interpretation of the absorption spectra of NRs.

\begin{figure}[!htbp]
  \centering
  \begin{minipage}[t]{7.5cm}
    \begin{center}
      \includegraphics[width=7.5cm]{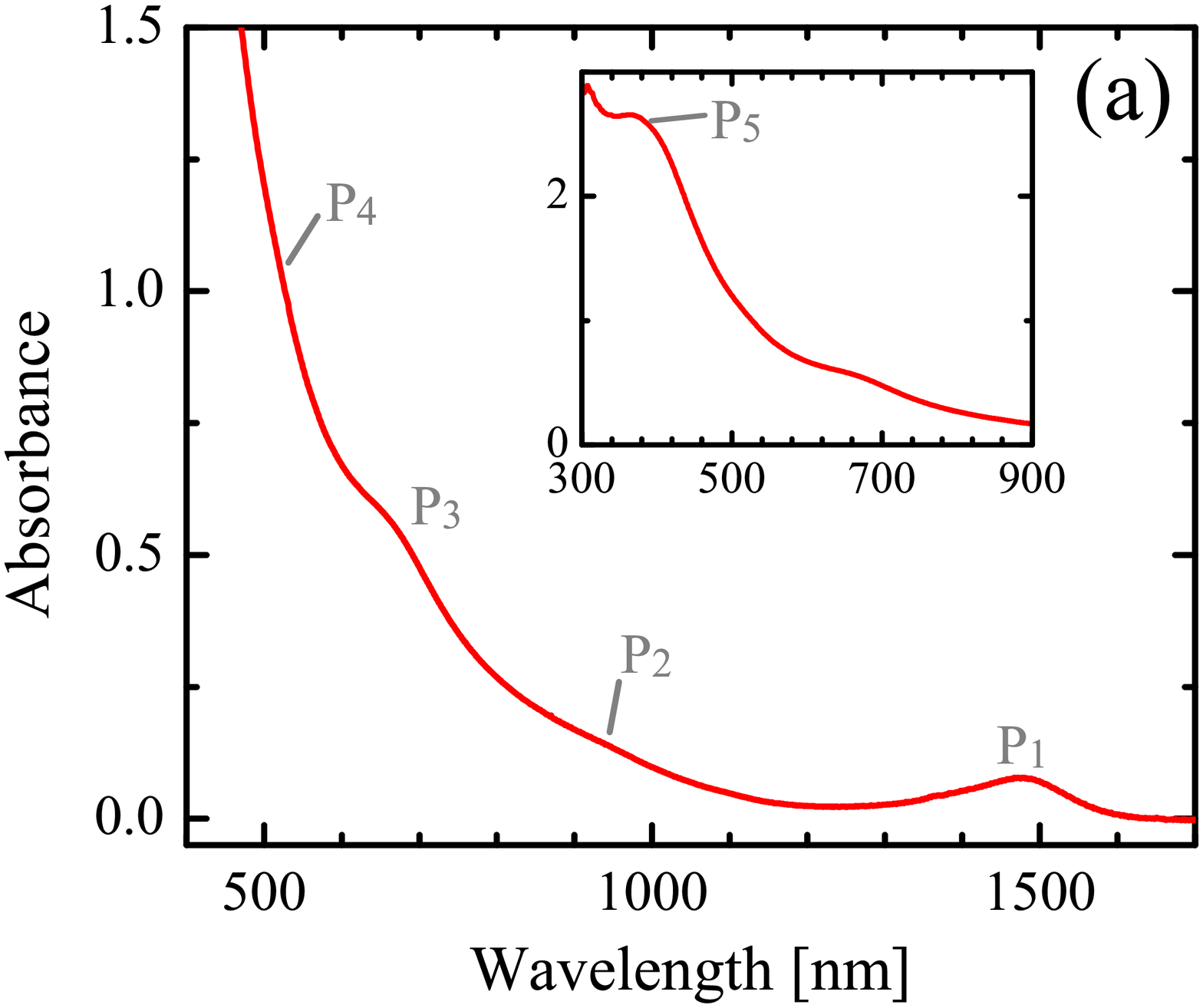}
    \end{center}
  \end{minipage}
  \begin{minipage}[t]{7.5cm}
    \begin{center}
      \includegraphics[width=7.5cm]{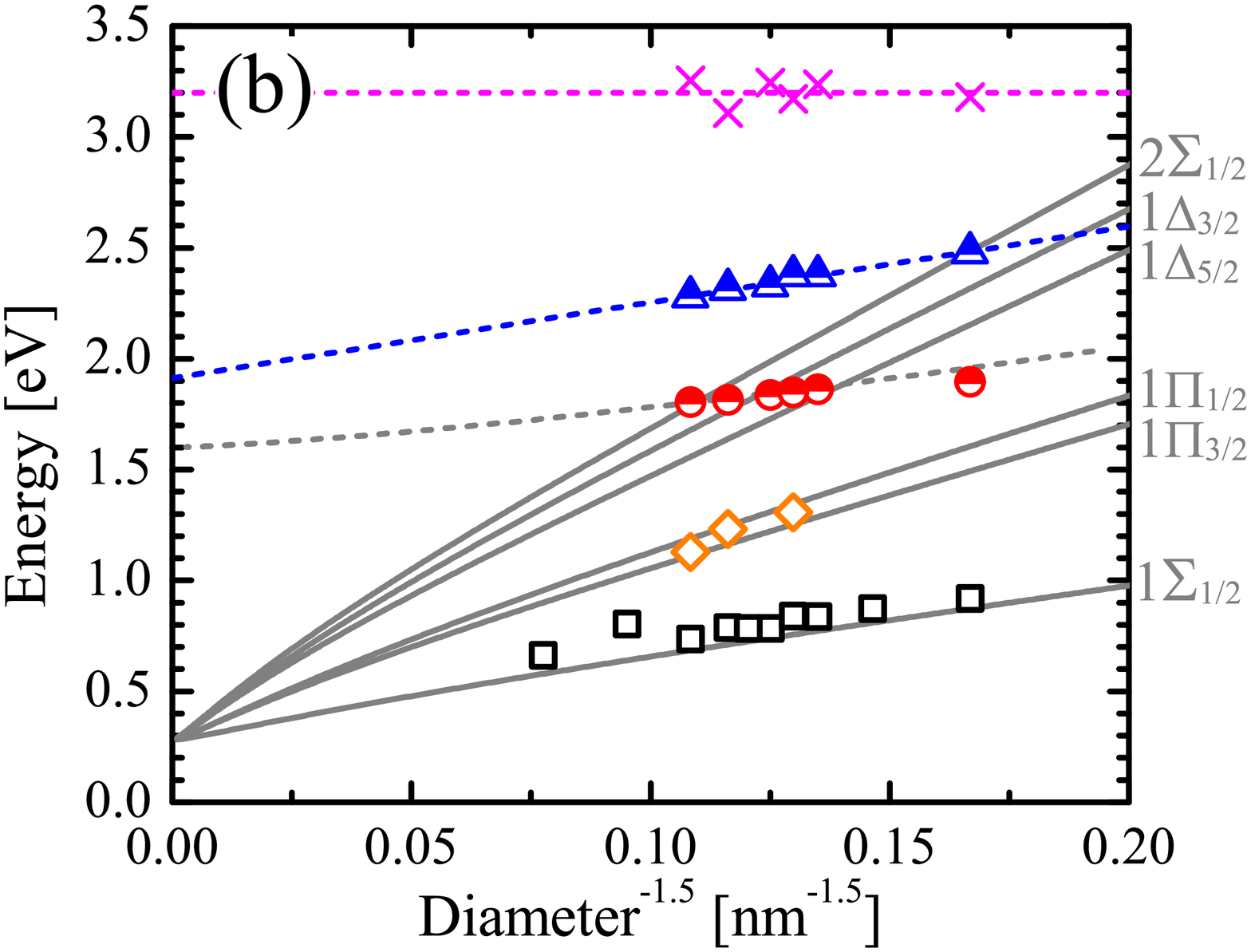}
    \end{center}
  \end{minipage}
  \caption{(a) Example absorption spectra of a 3.9 x 17 nm PbSe NR. Inset shows the same data, but on a scale where the 5th peak is visible. (b) Peaks in 2nd derivative spectrum as a function of NR diameter (symbols), calculated allowed transitions (grey lines), simple parabolic effective mass calculation around the $\Sigma$-point (dashed grey line), and linear fits (colored dashed lines.)}
  \label{expenergygap}
\end{figure}

We performed second--derivative analysis on the absorption spectra to determine the transition energies accurately. To avoid the problems inherent in this method \cite{Moreels_2ndderiv}, only the peaks in the second-derivative spectra that correspond to obviously-visible peaks in the measured spectra were used.  NRs produced by our first syntheses showed instability in solution and would slightly aggregate during the absorption measurement. This adds a moderate scattering background, so only the absorption peak location is recorded for these samples. NRs synthesized more recently are more stable, and at least four peaks can be discerned, with an additional peak in the three samples with narrowest size distribution. Fig. \ref{expenergygap}a has an example measured spectrum of a 3.9 nm diameter x 17 nm length PbSe NR that shows all five peaks, and the locations of all measurable peaks from all samples are shown in Fig. \ref{expenergygap}b. The measured peaks are plotted {\em vs.} $D^{-3/2}$ following the similar graph in Ref. \inlinecite{koole}. This power of the diameter is chosen to make the trend linear over the measured range, allowing rough extrapolation to bulk as $D^{-3/2} \to 0$. In this manner, the peaks originating from the L-point and $\Sigma$-point are easily distinguished.

Quantitative theoretical description of the size-dependent absorption spectra of PbSe NRs shown in Fig. \ref{expenergygap} requires a set of 6 room temperature energy band parameters for this semiconductor: $m_t^\pm$, $m_l^\pm$, and $P_{t,l}^2$. The parameters extracted from low temperature cyclotron resonance and interband magnetooptical experiments in bulk PbSe \cite{Pascher} describe quite well the average two-dimensional effective mass of electrons and holes at the bottom of the conduction band and the top of the valence band, respectively. The fitting procedure that gives this set is not sensitive, however, to the separation of $1/m_{l,t}$  and the $2P^2_{l,t}/m_0^2E_g$ terms, and describes well only the sum of these terms, because the all measurements are conducted a the narrow energy range comparable with the PbSe energy gap. This procedure is also not very sensitive to the anisotropy of the carrier energy spectra, because a magnetic field averages out the 2D motion of electrons and holes. On the other hand, in order to predict nanocrystal energy levels quantitatively, both the separation of components of the effective masses and the band anisotropy are crucial. Finally, the energy band parameters are expected to be temperature dependent. Thus, we conclude that parameters inferred from cyclotron resonance and magneto-optical measurements might not describe the energy spectra of NRs and NCs measured at room temperature.

In principle, spatial confinement of carriers in nanostructures provides a more-sensitive way to determine the energy band parameters, due to the large modification of the energy spectra of confined carriers. With this motivation, we used the previously-measured absorption spectra of PbSe NCs in Refs. \inlinecite{murray,moreels,wehrenburg,steckel,yu,koole} and extracted room-temperature band parameters using a global fitting procedure. Importantly, this new set of parameters not only quantitatively describes the low-energy transitions of PbSe NCs, but may also help resolve the long-standing controversy over the symmetry of the second peak in the NC absorption spectra (see Appendix \ref{bandchoice}). These band parameters (Table \ref{BandParamsTable}) are used in all graphs presented in this work.

\begin{table}[ht]
\centering
\begin{tabular}{||c|c|c||c|c|c||}
\hhline{|t:===:t:===:t|}
Name & Ref. \inlinecite{Pascher} & Best Fit  & Aniso. ratio & Ref. \inlinecite{Pascher} & Best Fit \\
\hhline{|:=:=:=::=:=:=:|}
$m_t^+/m_0$ & 0.29 & 0.59  & $m_l^+/m_t^+$ & 1.28 & 1.6 \\
\hhline{||-|-|-||-|-|-||}
$m_t^-/m_0$ & 0.27 & 0.79  & $m_l^-/m_t^-$ & 3.53 & 1.6 \\
\hhline{||-|-|-||-|-|-||}
$2 P_t^2 / m_0$ & 3.6 (eV) & 4.25 (eV) & $P_t^2 / P_l^2$ & 1.82 & 3.0\\
\hhline{|b:===:b:===:b|}
\hline
\end{tabular}
\label{BandParamsTable}
\caption{Energy band parameters that provide the best fits to the room temperature data from PbSe NCs. The left columns show the transverse band components, while the right columns show the ratio of transverse to longitudinal components.}
\end{table}

The theoretical size dependence of the optical transitions in PbSe NRs is calculated within our 4 band model and shown in Fig. \ref{expenergygap}b by solid lines. The lowest two transitions agree well with the theory. The third predicted transition is not observed, possibly owing to its proximity to other strong transitions in our NR samples. The third and fourth peaks are strong transitions that do not appear to be associated with the L-point. Their energies extrapolate back to the $\Sigma$-point energy. The third peak is fit well by the same parabolic band model used to model spherical PbSe NCs, and thus we assign this transition to the lowest-energy excitonic state at the $\Sigma$ point. This line was calculated for both spheres and rods with $m_\Sigma^e = m_\Sigma^h = 0.45 m_0$ and $E_g(\Sigma) = 1.65$ eV.  Without more-detailed knowledge of the band structure there, we cannot predict the excited states with any accuracy. Thus, the identity of the fourth transition cannot be determined, but as the energies approach the same 1.65-eV bulk value, it is reasonable to tentatively attribute it to a higher-energy exciton from the $\Sigma$ point. Finally, the fifth peak was perhaps the strongest in the absorption spectra, but showed no size dependence. We tentatively ascribe this to a metal-complex transition on the surface of the nanocrystal based on its proximity to absorption peaks of Pb(II) complexes \cite{divalent_lead}. The identities of these transitions are summarized in Table \ref{TransTable}.

\begin{table}[ht]
\centering
\begin{tabular}{|c|c|c|c|}
\hline Label & Assigned Transitions \\
\hline $P_1$ & $1 \Sigma_{1/2}^h \to 1 \Sigma_{1/2}^e$ \\
\hline $P_2$ & $1 \Pi_{3/2}^h \to 1 \Pi_{3/2}^e$ and  $1 \Pi_{1/2}^h \to 1 \Pi_{1/2}^e$ \\
\hline $P_3$ & $\Sigma$--point ground state \\
\hline $P_4$ & $\Sigma$--point excited state (?) \\
\hline $P_5$ & Surface metal complex mode\\
\hline
\end{tabular}
\label{TransTable}
\caption{Transitions observed in the absorption spectra of PbSe NRs.}
\end{table}

The fluoresence spectra and decays (Fig. \ref{QDvsNRgraph}b) are nearly identical for NCs and NRs, with a slightly larger Stokes shift in the NRs along with a slightly broader peak. The ensemble quantum yield of the nanorods is around 15\%, around half that of the NCs. This might indicate that the radiate lifetime of the rods is longer than the that of the NCs, but it is also possible that the QY reflects an ensemble with 15\% emitting and 85\% non-emitting rods.

Two effects would be expected to modify the radiative lifetime in nanorods. First, because the radiative lifetime is inversely proportional to the oscillator strength, the increased electron--hole correlation in NRs should decrease the lifetime compared to NCs. Second, the effect of screening is reduced in NRs, which is believed to be the cause of the long lifetime in PbSe NCs\cite{wehrenburg}. Approximating the NR as a dielectric prolate spheroid, the screening will substantially decrease along the rod axis, while slightly increasing along the other two axes, with an overall effect of a reduction in screening of the lifetime. Compared to a spherical NC of the same diameter, the larger oscillator strength and the reduced screening should each produce about a factor of 3 reduction in lifetime in NRs with typical aspect ratios.  Together this amounts to almost an order of magnitude reduction, and should be measurable even considering other sample--related uncertainties. However, the measured lifetime (Fig. \ref{QDvsNRgraph}b) is nearly identical in NCs and NRs. This discrepancy is not understood.  It might be explained by a dark ground exciton state that controls the photoluminescence decay in PbSe NRs and NCs, with the same activation mechanism in both structures. To be thorough, the nonradiative rate must be determined, and completing this along with exploring this phenomenon is a topic of future work.

\section{Discussion and Conclusions}

Our model of the electronic structure of lead--salt NRs is based on the 4 band ${\bm k} \cdot {\bm p}$ Hamiltonian suggested in  Ref. \inlinecite{mitchell_wallis}, using the standard boundary condition of a vanishing envelope wave function at the NR surface. All calculations are conducted within a cylindrical approximation. To use this model for description of various properties of NRs or NWs, one needs to know a set of the 6 temperature-dependent band parameters that describe a specific bulk lead--salt semiconductor. For the PbSe NRs studied in this paper, we extracted the set of room-temperature parameters from analysis of the size-dependence of previously-measured room temperature absorption spectra of spherical PbSe NCs.

The most significant conclusion of this work is that the fundamental excitations in PbSe NRs are one-dimensional excitons under each pair of optically coupled electron--hole subbands. The binding energy of the ground exciton state, which accumulates the most oscillator strength, increases with decreasing NR thickness and reaches 400\,meV in the narrowest rods. Surprisingly, the large binding energy of the exciton is almost exactly compensated  by the self--interaction of electrons and holes with their own images, which makes the energies of the optical transitions nearly independent of the solvent dielectric constant. Although the finite length of NRs affects the spacing between excited exciton states, it has a negligible effect on the energy of the exciton ground states.

With the set of PbSe band parameters extracted from spherical NC absorption spectra (Table \ref{BandParamsTable}), the model presented here describes the absorption spectra of PbSe NRs, and potentially resolves some troublesome aspects of ${\bm k} \cdot {\bm p}$ theory of spherical PbSe NCs. The energy of the optical transitions to the exciton ground states calculated within a cylindrical approximation match the two lowest-energy transitions observed experimentally. Although the effect of anisotropy in important for description of the absorption in spherical PbSe NCs, it is diminished in NRs (see Appendix \ref{pertappendix} \& \ref{bandchoice}), and the energy of the first two transitions is unaffected by it.

The absorption spectra of PbSe NRs have another remarkable feature.  The size dependence of the third and fourth absorption peaks is strong evidence that they originate from the $\Sigma$ point of the Brillouin zone.  Similar states connected with the $\Sigma$ point were observed previously in the absorption spectra \cite{koole} and in the hot carrier dynamics \cite{Cho_bulklike} of spherical PbSe NCs. These observations provide clear experimental evidence that even in the smallest nanostructures, wave functions from distinct critical points ($L$ and $\Sigma$, in this particular case) are not mixed if both their corresponding conduction band minima and valence band maxima are well-separated energetically. This provides strong justification for the applicability of our multiband effective mass approximation in such small nanostructures. A large energetic separation of $L$ and $\Sigma$ band edges is supported theoretically by recent \emph{ab initio} calculations \cite{DFTGW}, which for PbSe predict larger than 500 meV energy separation for these extrema, in both the valence and conduction bands, in contradiction with similar earlier calculations, which placed the separation in the valence band closer to 150\,meV \cite{an_zunger}.

The predicted strong increase in electron--hole Coulomb interaction in PbSe NWs should have major implications for other properties. This enhancement should increase the rate of the nonradiative Auger recombination as well as the rate of the inverse process, impact ionization. A high rate of impact ionization or efficient multiple exciton generation, combined with good conductivity that might be expected in PbSe NWs, suggests that these structures may be promising for photovoltaic applications.

To summarize, we have developed a theory that describes both the energy spectra of individual electrons and holes and the absorption spectra of lead--salt NWs and NRs.  Calculations show that even though spatial and dielectric confinement dramatically increase the exciton binding energy, the absorption spectra of PbSe NWs and NRs are practically unaffected, which should lead to insensitivity of these spectra to the surrounding media. The size dependence of lowest absorption peaks measured in PbSe NRs is very well described by the developed theory.  It should be straightforward to apply this model to PbS and PbTe NRs.

\appendix
\section{Effect of anisotropy on the nanowire energy spectra \label{pertappendix}}

\begin{figure}[!htbp]
  \centering
  \begin{minipage}[t]{7.5cm}
    \begin{center}
      \includegraphics[width=7.5cm]{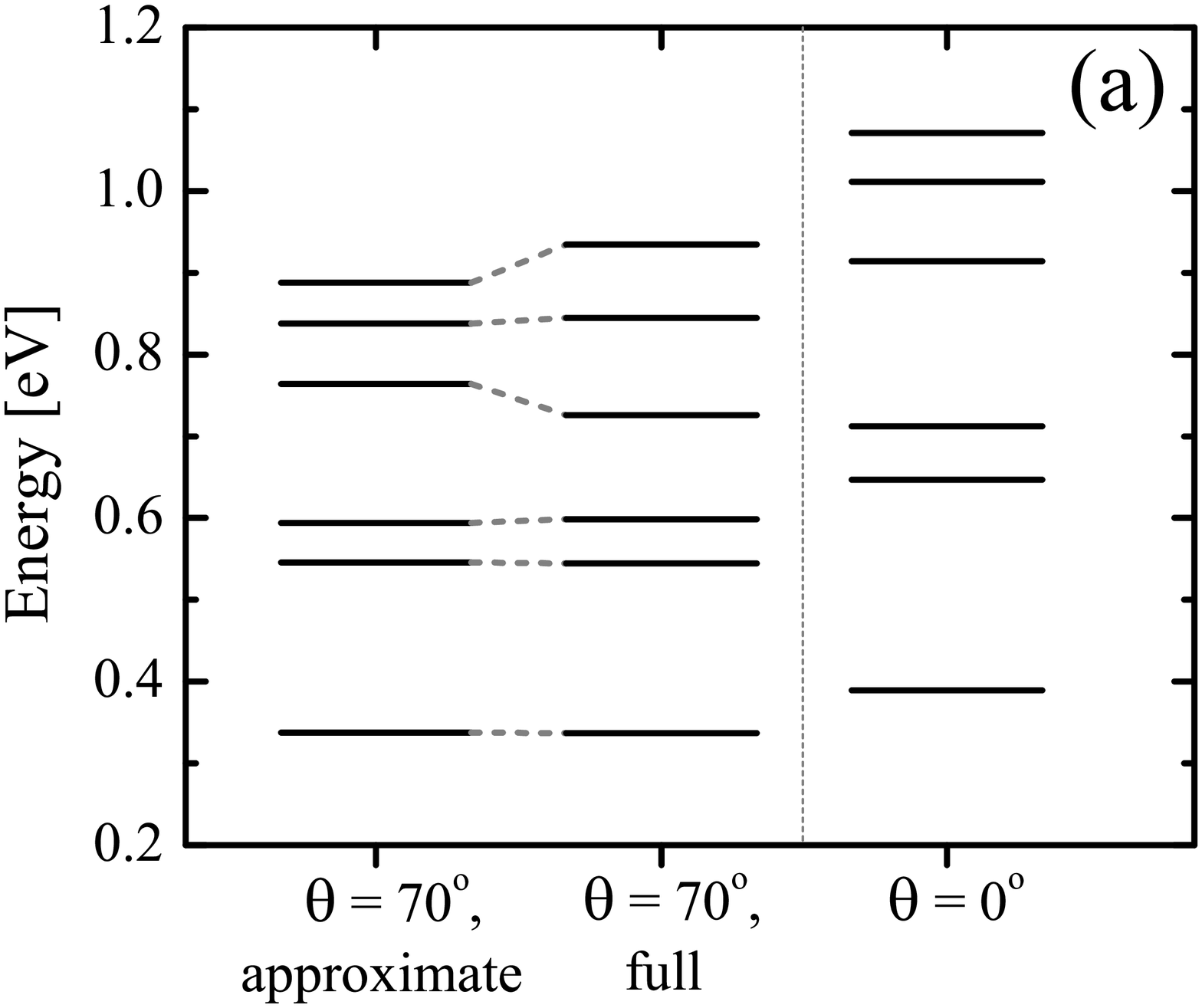}
    \end{center}
  \end{minipage}
  \begin{minipage}[t]{7.5cm}
    \begin{center}
      \includegraphics[width=7.5cm]{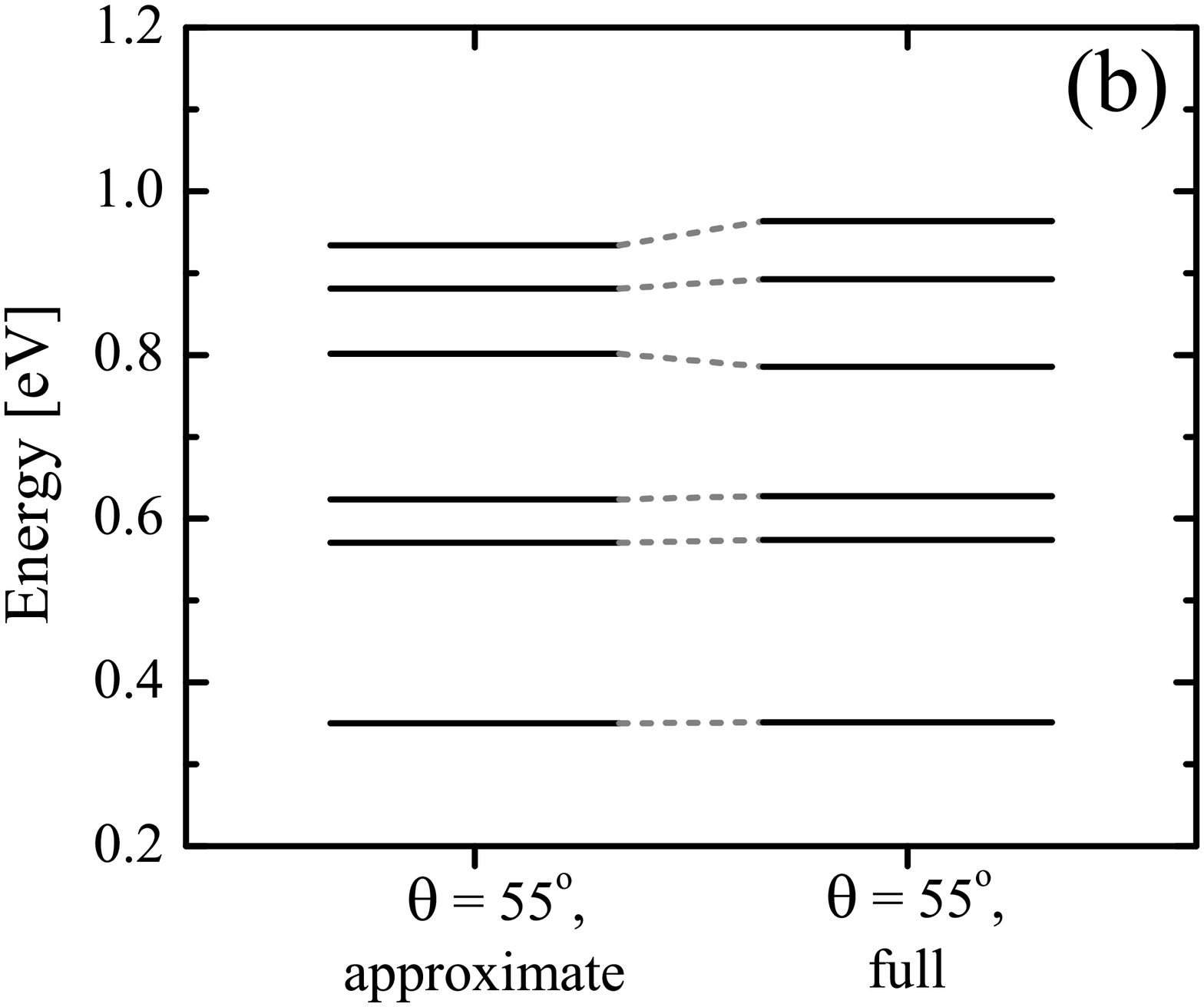}
    \end{center}
  \end{minipage}
  \caption{Effect of the energy spectrum anisotrpy on the energy of the 1D subband bottom in a 4 nm PbSe NW grown along the (a) $\langle 111 \rangle$ and (b) $\langle 100 \rangle$ crystal axes. The ``approximate'' calculations are conducted within the cylindrical approximation, which gives Eq. \eqref{det} for the energy levels. The ``full'' calculations are performed as described in the text. The energy levels are labeled by the angle between the L-point and the rod growth axis. Note that the $\theta = 0$ energy levels do not require perturbation, as $H_{an}=0$ for that angle.}
  \label{pertfig}
\end{figure}

The cylindrically symmetric  Hamiltonian in Eq. \eqref{hcyl}  can be derived from the full Hamiltonian in Eq. \eqref{hfull} by transformation to the new coordinate system connected with NW direction. The full Hamiltonian  is defined with respect to a crystallographic direction of the Brillouin zone, where the z--axis is pointed towards one of the L--points, and we will call this coordinate system the primed system, $\{x',y',z'\}$. We need to express Eq. \eqref{hfull} in the new coordinate system where the z--axis is directed along the rod axis, called the unprimed system, $\{x,y,z\}$. To do this, we  use a coordinate rotation, and  define the x--axis such that the rotation occurs in the x--z plane. In the rotation, vector quantities, such as $\hat{\bf p}$ or $\hat{\boldsymbol{\sigma}}$  are transformed using the rotation matrix, $\hat{\bf p}' = R(\theta) \hat{\bf p}$, with R defined as
\be
R(\theta) = \begin{pmatrix}
\cos \theta & 0 & -\sin\theta \\
0 & 1 & 0 \\
\sin \theta & 0 & \cos\theta
\end{pmatrix} ~.
\ee
This transformation expresses the squared momenta in Eq. \eqref{hfull} as:
\bea
\hat{p}_x'^2 & = & \cos^2\theta \hat{p}_x^2 - \sin2\theta \hat{p}_x \hat{p}_z + \sin^2\theta \hat{p}_z^2 \\
\hat{p}_z'^2 & = & \sin^2\theta \hat{p}_x^2 + \sin 2 \theta \hat{p}_x \hat{p}_z + \cos^2\theta \hat{p}_z^2~.
\eea
and the diagonal and off-diagonal elements of the matrix of Hamiltonian in Eq. \eqref{hfull} in new coordinate system as:
\bea
\frac{1}{m_t} (\hat{p}_x'^2 + \hat{p}_y'^2 ) + \frac{1}{m_l} \hat{p}_z'^2 & = & \left( \frac{\cos^2 \theta}{m_t} + \frac{\sin^2 \theta}{m_l} \right) \hat{p}_x^2 + \frac{1}{m_t} \hat{p}_y^2 +\nonumber \\
&  & + \left( \frac{\sin^2 \theta}{m_t} + \frac{\cos^2 \theta}{m_l} \right) \hat{p}_z^2 + \nonumber\\
&  & +  \sin 2 \theta \left( \frac{1}{m_l} - \frac{1}{m_t} \right) \hat{p}_x \hat{p}_z \\
P_t \sigma_x' \hat{p}_x' + P_t \sigma_y' \hat{p}_y' + P_l \sigma_z' \hat{p}_z' & = & (P_t \cos^2\theta + P_l \sin^2 \theta) \sigma_x \hat{p}_x + P_t \sigma_y \hat{p}_y + \nonumber \\
&  & + (P_t \sin^2 \theta + P_l \cos^2 \theta) \sigma_z \hat{p}_z + \nonumber \\
&  & + \frac{1}{2} \sin 2 \theta (P_l - P_t)(\sigma_z \hat{p}_x + \sigma_x \hat{p}_z)
\eea
Notice that neither elements are cylindrically symmetric in the new coordinates. To enforce this symmetry, we rewrite these expressions in a form that separates a cylindrically symmetrical part, formally: $a \hat{O}_x + b \hat{O}_y = (1/2) (a+b) (\hat{O}_x + \hat{O}_y) + (1/2) (a-b) (\hat{O}_x-\hat{O}_y)$. The first term, which has cylindrical symmetry, is used in the zero-th order Hamiltonian, and the second term creates the asymmetric perturbation. This procedure produces the Hamiltonian in Eq. \eqref{hcyl}, along with the perturbation matrix
\be
\hat{H}_{an} =
\begin{pmatrix}
\begin{matrix}\frac{1}{2} ~ \hat{U} \left(\frac{1}{m_l^-} - \frac{1}{m_t^-}\right) \times \\
\times \left( \frac{1}{2} \sin^2 \theta (\hat{p}_x^2 - \hat{p}_y^2) + \sin2 \theta \hat{p}_x \hat{p}_z \right) \\
~
\end{matrix} & \quad &
\begin{matrix}\frac{1}{2 m}(P_l - P_t) \big\{ \sin^2 \theta (\hat{\sigma}_x \hat{p}_x - \hat{\sigma}_y \hat{p}_y) + \\
+ \sin 2 \theta (\hat{\sigma}_z \hat{p}_x + \hat{\sigma}_x \hat{p}_z ) \big\} \\
~
\end{matrix} \\
\begin{matrix}\frac{1}{2 m}(P_l - P_t) \big\{  \sin^2 \theta (\hat{\sigma}_x \hat{p}_x - \hat{\sigma}_y \hat{p}_y) + \\
+ \sin 2 \theta (\hat{\sigma}_z \hat{p}_x + \hat{\sigma}_x \hat{p}_z ) \big\}
\end{matrix} & \quad &
\begin{matrix}-\frac{1}{2} ~ \hat{U} \left(\frac{1}{m_l^+} - \frac{1}{m_t^+}\right) \times \\
\times \left( \frac{1}{2} \sin^2 \theta (\hat{p}_x^2 - \hat{p}_y^2) + \sin2 \theta \hat{p}_x \hat{p}_z \right)
\end{matrix}
\end{pmatrix} \label{Han}
\ee
We study the effect of anisotropy described by Eq. \eqref{Han} on the energy spectrum of electrons and holes. Figure \ref{pertfig} compares the energy of the lowest electron levels in a 4 nm PbSe NW calculated within the cylindrical approximation and with complete numerical inclusion of the anisotropy. The anisotropy was taken into account by diagonalizing the matrix elements of $H_{an}$ in the space of the highest 20 valence and lowest 20 conduction states (that is, including the highest ten and lowest ten doubly degenerate electron and hole levels.) One can see in Fig. \ref{pertfig} that the anisotropy in PbSe splits the nearly degenerate energy levels, whose radial or angular quantum momentum numbers differ by one in radial or angular quantum momentum numbers, while necessarily leaving the Kramer's degeneracy unbroken. The splitting should broaden the energy levels without an overall shift in the level position.

\section{Calculations of the one dimensional Coulomb potential \label{coulappendix}}

Calculation of the one dimensional  Coulomb potential in Eq. \eqref{Vavg} and self interaction energy in Eq. \eqref{Eself} can be greatly simplified by initial averaging over angular variables. For the $U_1$ term of Eq. \eqref{Vavg} the angular integration results
\be
\langle U_1 \rangle(z) = \int_0^R d\rho_e \rho_e   \int_0^R d\rho_h \rho_h |\Psi_e|^2 |\Psi_h|^2 V_1(\rho_e, \rho_h, z)~,
\ee
where
\be
V_1(\rho_e, \rho_h, z) = - 4 \pi \frac{e^2}{\kappa_s \sqrt{\rho_e \rho_h}} Q_{-1/2}\left(\frac{z^2 + \rho_e^2 + \rho_h^2}{2 \rho_e \rho_h}\right)
\ee
and $Q_n$ is the Legendre function of the second kind. The two remaining radial integrals are evaluated numerically.

For the second term in Eq. \eqref{Vavg}, $U_2$, the angular integrals vanish unless $m=0$ leaving only this term from the sum. This results in the following expression for $\langle U_2 \rangle(z)$:
\begin{eqnarray}
\langle U_2 \rangle(z) & = & - 8 \pi \frac{e^2}{\kappa_s}\int_0^\infty d u \,  \frac{(\kappa_s - \kappa_m) K_0(R u) K_1(R u) \cos(u z)}{\kappa_s I_1(R u) K_0(R u) + \kappa_m I_0(R u) K_1(R u)} \times \nonumber \\
& \times & \underbrace{\left( \int_0^R d\rho_e \, \rho_e |\Psi_e|^2 I_0 (u \rho_e) \right)}_{i_e(u)} \underbrace{\left( \int_0^R d\rho_h \, \rho_h |\Psi_h|^2 I_0 (u \rho_h) \right)}_{i_h(u)}~. \label{U2ints}
\end{eqnarray}
To calculate the integrals $i_e$ and $i_h$ in Eq. \eqref{U2ints}, we approximate the squared wavefunctions as a short sum of the form $|\Psi_e|^2 = \sum_{n=1}^{N} A_n (1 - \rho_e^{2 n})$, with $N \approx 8$. Even with so few terms, the maximum relative error is typically $<10^{-7}$. This allows us to solve these two integrals analytically:
\begin{eqnarray}
i_e(u) & = & \sum_{n=1}^N A_n \int_0^R d\rho_e \, \rho_e ( 1 - \rho_e^{2 n} ) I_0 (u \rho_e) \nonumber \\
& = & \sum_{n=1}^N A_n \left( \frac{R I_1(u)}{u} - \frac{R^{2 + 2 n} \,_1F_2\left(1 + n ; 1 , 2 + n ; R^2 u^2 / 4 \right)}{2 + 2 n} \right)~, \label{polyapprox}
\end{eqnarray}
where $_pF_q$ is the generalized hypergeometric function. The remaining integral over $u$ in Eq. \eqref{U2ints} is performed numerically.

Lastly, the two self interaction terms in  Eq. \eqref{Eself}, $U_e$ and $U_h$, after angular integrations are reduced to
\begin{eqnarray}
& &\langle U_{e,h}\rangle  = \frac{2 e^2}{\kappa_s} \sum_{m=0}^\infty \int_0^\infty du \left( \int_0^R d\rho_{e,h} \, \rho_{e,h} |\Psi_{e,h}|^2   I_m^2(u \rho_{e,h}) \right) \times \nonumber \\
& & \times  \frac{(\kappa_s - \kappa_m)K_m(R u)(K_{m-1}(R u) + K_{m+1}(R u))(2-\delta_{m 0})}{\kappa_s K_m(R u) (I_{m-1}(R u) + I_{m+1}(R u)) + \kappa_m I_m(R u)(K_{m-1}(R u) + K_{m+1}(R u))}~.
\label{sien}
\end{eqnarray}
The two dimensional integrals in Eq. \eqref{sien} was taken numerically. It is summed over only the first $\approx$20 values of $m$, as the sum converges rapidly.

\section{Numerical calculation of the exciton binding in PbSe nanorods \label{comappendix}}

Our analytic model makes the assumption that the 1D exciton is only weakly confined along the NR axis. In this case the finite length of the NR affects only the exciton center of mass motion. To verify this assumption, the 1D Hamiltonian was numerically diagonalized, while treating both binding and confinement exactly. As an orthogonal basis for this diagonalization we used a sufficiently large set of electron and hole plane waves that satisfied the single particle boundary conditions. The 1D exciton wavefunction
in this basis set can be written as:
\be
\Psi_{1D} = \sum_{n_e = 1}^{N_e} \sum_{n_h = 1}^{N_h} A_{n_e, n_h} \frac{2}{L} \sin\left(\frac{n_e \pi z_e}{L}\right) \sin\left(\frac{n_h \pi z_h}{L}\right)~
\ee
where  $A_{n_e, n_h}$ are the numerical coefficients.

The kinetic energy is diagonal in this basis, and matrix elements of Eq. \eqref{Ueff} can be evaluated analytically. Calculation time was dominated by evaluation of these matrix elements and scaled as O($N_e N_h$). For $N_e = N_h \approx 30$, calculations were sufficiently converged for the lowest few dozen states, and required roughly one minute of computation time on a desktop computer.

\begin{figure}[htbp]
    \centering
    \includegraphics[width=14cm]{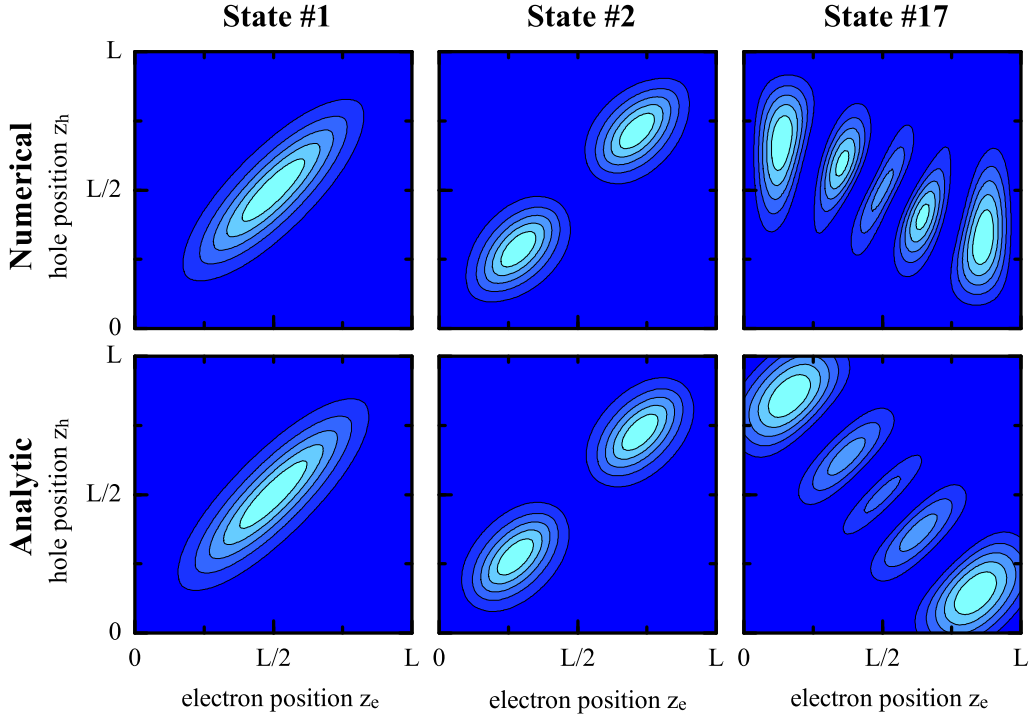}
    \caption{Comparison of the numerically and analytically calculated 1D exciton wavefunctions $|\Psi_{1D}|^2$. Each subplot has axes $z_e$ and $z_h$ ranging along the length of the nanorod from $0$ to $L$. The two lowest energy states and the 17th state are shown.}
    \label{statesfig}
\end{figure}

Fig. \ref{statesfig} shows the square of 1D wavefunctions, $|\Psi_{1D}|^2$, calculated  both the numerically and analytically  as a function of $z_e$ and $z_h$. For the lowest two exciton states $|\Psi_{1D}|^2$ shows good agreement between the numerical model and the analytical calculation. This is because the electron and hole are strongly localized around each other and do not feel the effects of confinement at the edges of box. As a result, the wavefunction orients along the coordinates associated with Coulomb binding, $z$ and $z_\text{com}$, roughly along the graph diagonals. On the other hand, by the 17th excited state, also shown in Fig. \ref{statesfig}, the numerical and analytical calculations disagree greatly. This is because the higher kinetic energy of this state causes the wavefunction to reach the edges of the box and feel confined. And, as a result, it begins to orient along the box coordinates, $z_e$ and $z_h$, associated with confinement. %To better model this state analytically, one should treat binding as a perturbation instead.
In general, our analytic model shows good agreement for the lowest $\approx 10$ states for each pair of nanowire bands.

\section{Choice of the room temperature band parameters \label{bandchoice}}

The absence of reliable room temperature energy band parameters for bulk PbSe has lead to several problems in the quantitative description of spherical PbSe NC electronic properties within effective mass theory, and as a result, to some controversy on their electronic structure \cite{an_zunger, allan_delerue, kilina_prezhdo, franceschetti_dft, koole}. As has been noted \cite{lipovskii, allan_delerue}, effective mass theory significantly overestimates the energy gap in PbSe NCs (though not in PbS.) In addition, the nature of the 2nd optical transition is still a source of debate \cite{du, peterson, franceschetti_2nd, trinh}, whether it is of symmetry type S-P or P-P. Considering the body of experimental evidence, the explanation put forward by Franceschetti \cite{franceschetti_2nd} seems to offer the simplest explanation of this controversy, that the electron and hole P states are split into $P_\perp$ and $P_\parallel$ states by the anisotropy of the bands, and the second transition is of type $P_\parallel$-$P_\parallel$. These two problematic aspects of experimental spectra of PbSe NCs for effective mass theory-- overestimation of the bandgap and the symmetry of the 2nd transition-- as well as the observation of several optical transitions in a wide range of energies can be used for extraction of a real set of the energy band parameters.

Although the extraction of the set of energy band parameters from room temperature absorption spectra is possible, it is likely that many sets of parameters will equally well fit the first few optical transitions. In order to increase the accuracy of the fit, we want to somehow incorporate the energy band parameters in low temperature experiments in bulk PbSe. So, the total band edge effective masses for electrons and holes at $T = 4$ K are held constant at the values from experiment \cite{Pascher}. In addition, to limit the degrees of freedom in the fit, the anisotropy of the far-band contributions to both the electron and hole are held equal. That is, $m_l^+ / m_t^+ = m_l^- / m_t^-$, even though their individual values will differ. With these constraints, a fit is performed using the body of literature data \cite{murray,moreels,wehrenburg,steckel,yu,koole} for the first transition, and the data from Koole \cite{koole} for the second and third transitions.

\begin{figure}[!htbp]
  \centering
  \begin{minipage}[t]{7.5cm}
    \begin{center}
      \includegraphics[width=7.5cm]{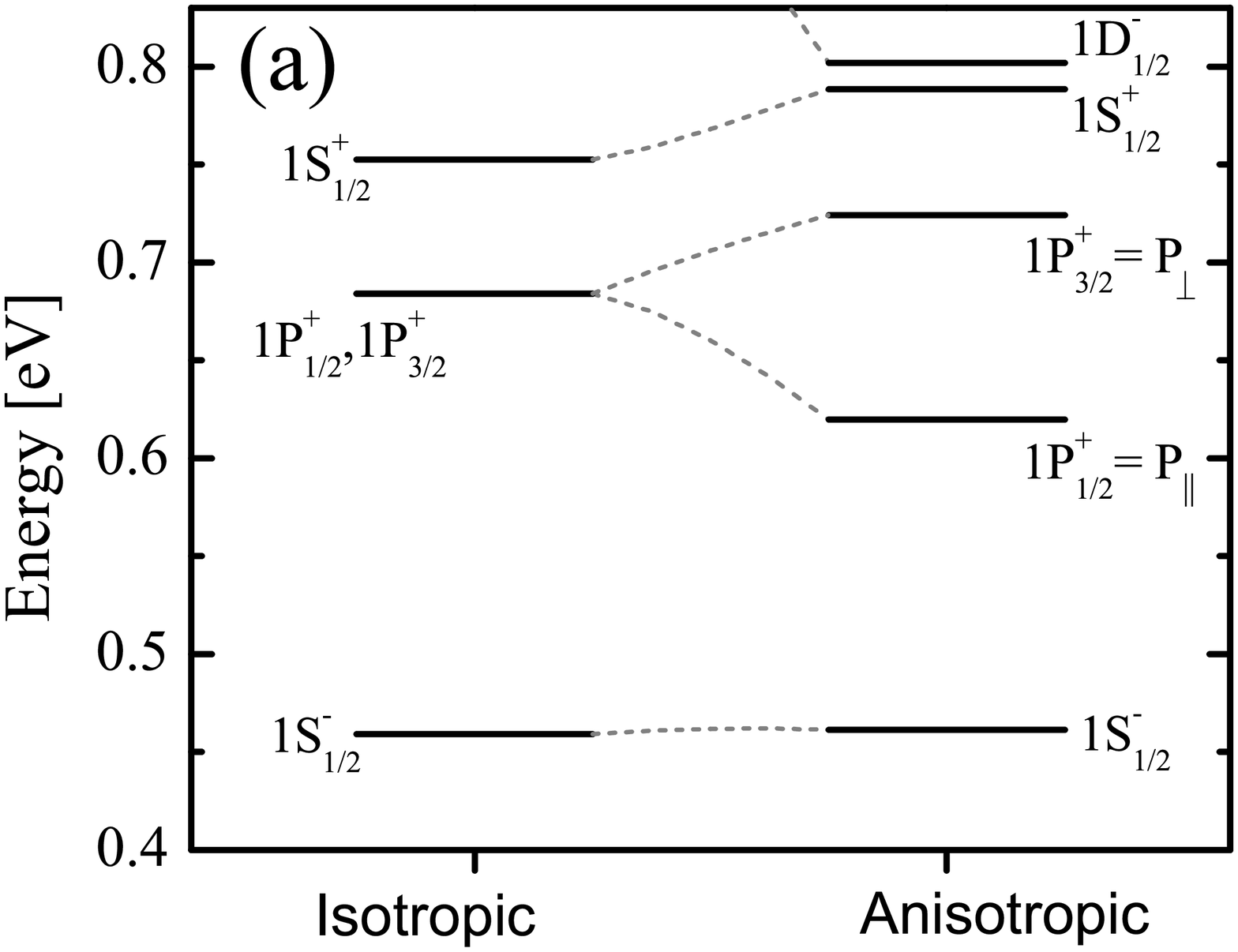}
    \end{center}
  \end{minipage}
  \begin{minipage}[t]{7.5cm}
    \begin{center}
      \includegraphics[width=7.5cm]{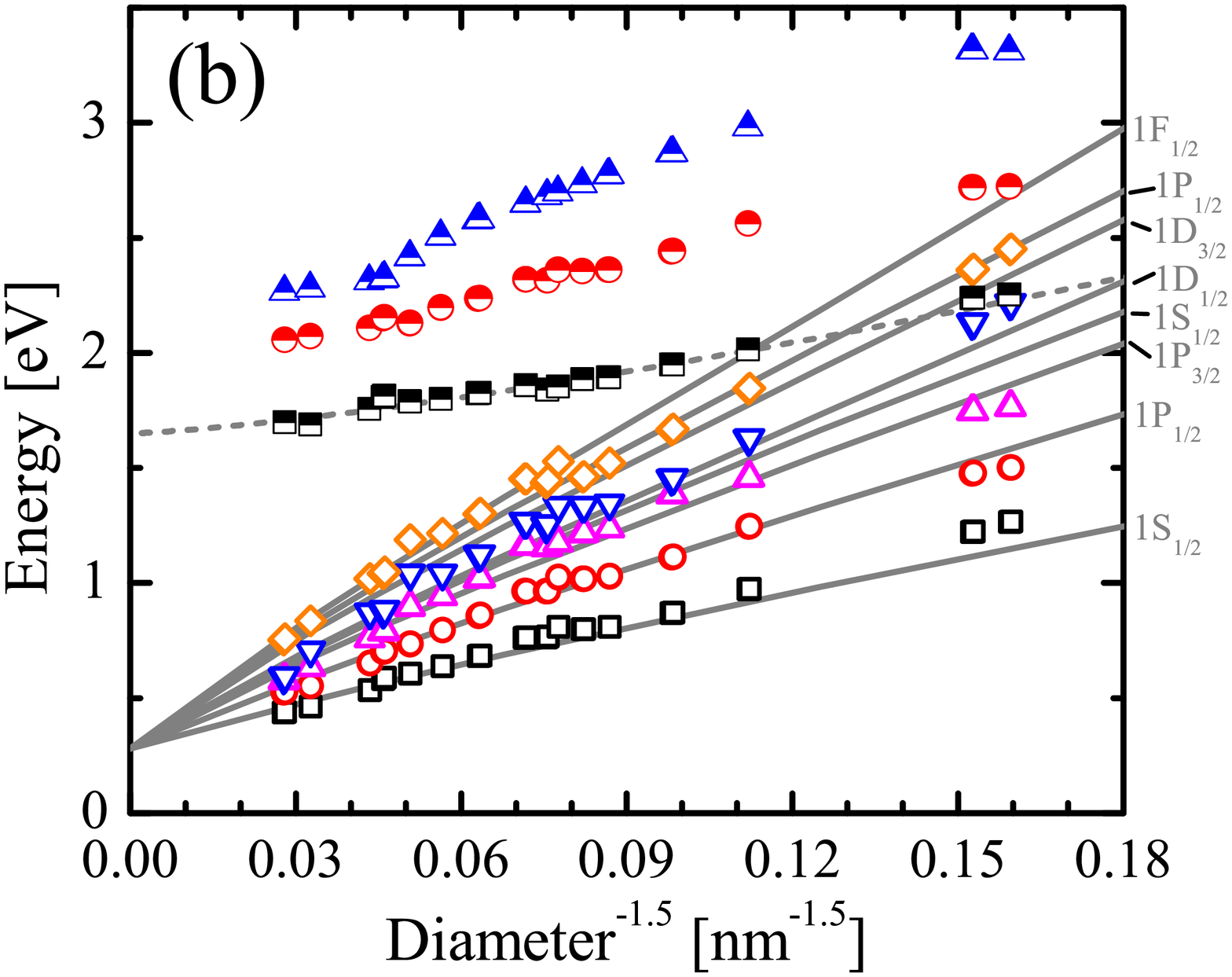}
    \end{center}
  \end{minipage}
  \caption{Calculations of the lowest electron levels in spherical PbSe NCs. (a) Splitting of the P state induced by the fully anisotropic Hamiltonian in a 4 nm radius NC. Anisotropic states are labeled by writing the state in the basis of isotropic states, and labeling it by the isotropic state with largest coefficient. (b) The size dependence of the transition energies in spherical PbSe NCs. Experimental data \cite{koole} are shown by symbols. The solid lines show the size dependence of optically allowed transitions calculated in a fully anisotropic effective mass model. The optically allowed transitions occur between the states of the same symmetry but opposite parity, and we label them by a symmetry type, which is common for both states. Open points indicate transitions originating from the L-point in the Brillouin zone, while half-open points are suggested to be from the $\Sigma$ point as in Ref. \inlinecite{koole}. The dashed line shows the size dependence of lowest confined level connected with the $\Sigma$ point of the Brillouin zone, calculated in a parabolic effective mass approximation as explained in the text.}
  \label{PbSeQDNewParams}
\end{figure}

The final set of room temperature parameters are shown in Table \ref{BandParamsTable} together with the set of low temperature parameters reported for bulk PbSe in Ref. \inlinecite{Pascher}. The transition energies calculated using these parameters are shown in Fig. \ref{PbSeQDNewParams}. The anisotropic effective mass calculations were performed using the method outlined in Ref. \inlinecite{tudury} and the results compared to the energies measured in Ref. \inlinecite{koole}, ignoring those points criticized in Ref. \inlinecite{Moreels_2ndderiv} as possibly being 2nd derivative artifacts.

\begin{acknowledgments}
A. C. Bartnik thanks Jun Yang for assistance with numerical optimizations and for enlightening discussions. This work was supported by the Cornell Center for Nanoscale Systems (CNS) through National Science Foundation Grant EEC-0646547, and in part by the Cornell Center for Materials Research (CCMR) with funding from the Materials Research Science and Engineering Center program of the National Science Foundation (cooperative agreement DMR 0520404). Al. L. Efros thanks the U.S. Office of Naval Research (ONR) and the U.S. Department of Energy (DOE). W.-k. Koh and C. B. Murray acknowledge financial support from the NSF through DMS-0935165. This research was partially supported by the Nano/Bio Interface Center through the National Science Foundation (NSEC DMR08-32802) with a seed grant that initiated the investigation of the synthesis  of the nanorods.
\end{acknowledgments}

\bibliography{Nanorod} % for the final draft

\end{document}